\documentclass[aps,floats]{revtex4}
\usepackage{amsmath,amssymb}
\usepackage{graphicx,epsfig}

\begin{document}
\bibliographystyle {plain}

\def\oppropto{\mathop{\propto}} 
\def\opsimeq{\mathop{\simeq}}
\def\opoverderline{\mathop{\overline}}
\def\operarrow{\mathop{\longrightarrow}}
\def\opsim{\mathop{\sim}}

\def\fig#1#2{\includegraphics[height=#1]{#2}}
\def\figx#1#2{\includegraphics[width=#1]{#2}}


\title{ Dynamical barriers of pure and random ferromagnetic Ising models \\
on fractal lattices  } 


 \author{ C\'ecile Monthus and Thomas Garel }
  \affiliation{ Institut de Physique Th\'{e}orique, CNRS and CEA Saclay,
 91191 Gif-sur-Yvette, France}

\begin{abstract}

We consider the stochastic dynamics of the pure and random ferromagnetic Ising model on the hierarchical diamond lattice of branching ratio $K$ with fractal dimension $d_f=(\ln (2K))/\ln 2$. We adapt the Real Space Renormalization procedure introduced in our previous work [C. Monthus and T. Garel, J. Stat. Mech. P02037 (2013)] to study the equilibrium time $t_{eq}(L)$ as a function of the system size $L$ near zero-temperature. For the pure Ising model, we obtain the behavior $t_{eq}(L) \sim L^{\alpha} e^{\beta 2J L^{d_s}} $ where $d_s=d_f-1$ is the interface dimension, and we compute the prefactor exponent $\alpha$. For the random ferromagnetic Ising model, we derive the renormalization rules for dynamical barriers $B_{eq}(L) \equiv (\ln t_{eq}/\beta)$ near zero temperature. For the fractal dimension $d_f=2$, we obtain that the dynamical barrier scales as $ B_{eq}(L)= c L+L^{1/2} u$ where $u$ is a Gaussian random variable of non-zero-mean. While the non-random term scaling as $L$ corresponds to the energy-cost of the creation of a system-size domain-wall, the fluctuation part scaling as $L^{1/2}$ characterizes the barriers for the motion of the system-size domain-wall after its creation. This scaling corresponds to the dynamical exponent $\psi=1/2$, in agreement with the conjecture $\psi=d_s/2$ proposed in [C. Monthus and T. Garel, J. Phys. A 41, 115002 (2008)]. In particular, it is clearly different from the droplet exponent $\theta \simeq 0.299$ involved in the statics of the random ferromagnet on the same lattice.

\end{abstract}

\maketitle

\section{ Introduction  }

Among real-space renormalization procedures  \cite{realspaceRG}, 
Migdal-Kadanoff block renormalizations \cite{MKRG} play a special role
because they can be considered in two ways, 
 either as approximate renormalization procedures on hypercubic lattices,
or as exact renormalization procedures on certain hierarchical lattices \cite{berker,hierarchical,derridapotts}.
Besides the study of pure models, these hierarchical lattices have been also much used to study the equilibrium of 
disordered classical spin models, 
such as the diluted Ising model \cite{diluted}, 
the random bond Potts model \cite{potts,us_potts,turban,auriac},
the random field Ising model \cite{caomachta,dayan}
and spin-glasses \cite{hierarchicalspinglass}. For spin-glasses, these  hierarchical lattices
have been also used to study dynamical properties \cite{hertz,ritort,maass,drossel}.
The equilibrium properties of disordered polymer models have also been considered,
in particular the  wetting on a disordered substrate \cite{Der_wett,Tang_Chate,us_poly}
and the directed polymer model
 \cite{Der_Gri,Coo_Der,Tim,roux,kardar,cao,tang,Muk_Bha,Bou_Sil,us_poly}.

In the present paper, we consider the pure and the random ferromagnetic Ising model
on the hierarchical diamond lattice in order to study its dynamical properties near zero temperature.
We adapt the real space renormalization procedure introduced in \cite{us_rgdyn} :
 using the standard mapping between the detailed-balance dynamics of classical Ising models
 and some quantum Hamiltonian,
we obtain the appropriate real-space renormalization rules for the quantum Hamiltonians
associated to single-spin-flip dynamics. This approach has been used previously
to characterize the dynamics of the random ferromagnetic chain 
 \cite{us_rgdyn}, of the pure Ising model on the Cayley tree  \cite{us_rgdyn}, 
of the random Ising model on the Cayley tree  \cite{us_rgdyntree}, and for
the hierarchical Dyson Ising model \cite{us_rgdyndyson}.

The paper is organized as follows.
In section \ref{sec_model}, we introduce the notations
for the stochastic dynamics of Ising models defined on the hierarchical diamond model.
In section \ref{sec_pure}, we analyze via real-space renormalization
the dynamics of the pure Ising model.
In section \ref{sec_random}, we study via real-space renormalization
the dynamics of the random ferromagnetic Ising model.
Our conclusions are summarized in section \ref{sec_conclusion}.
In Appendix \ref{sec_app}, we derive some renormalization formula that are used in the text.

\section{ Model and notations  }

\label{sec_model}

\subsection{ Hierarchical diamond lattice of branching ratio $K$}

\begin{figure}[htbp]
\includegraphics[height=6cm]{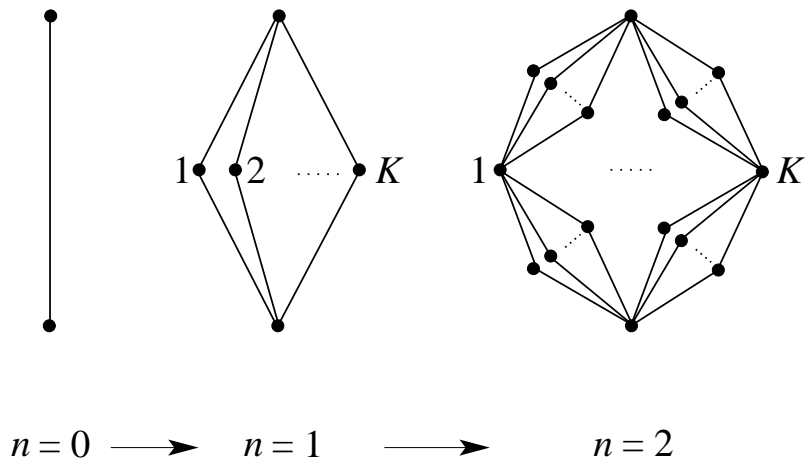}
\hspace{1cm}
\caption{ Hierarchical construction of the diamond lattice of
branching ratio $K$    }
\label{figdiamond}
\end{figure}

As shown on Fig. \ref{figdiamond}, the hierarchical diamond lattice of branching ratio $K$ 
is constructed recursively
from a single link called generation $n=0$  :
 generation $n=1$ consists of $K$ branches, each branch
 containing $2$ bonds in series ;
 generation $n=2$ is obtained by applying the same transformation
to each bond of the generation $n=1$.
At generation $n$, the length $L_n$ between the two extreme sites is 
\begin{eqnarray}
L_n=2^n
\label{lndiamond}
\end{eqnarray}
whereas the total number $B_n$ of bonds is 
\begin{eqnarray}
B_n=(2 K)^n=L_n^{d_f}
\label{bndiamond}
\end{eqnarray}
so that
\begin{eqnarray}
d_f= \frac{ \ln (2K)}{\ln 2}
\label{deffdiamond}
\end{eqnarray}
 represents the fractal dimension.

As recalled in the introduction, the equilibrium
of many statistical physics models have been studied on this lattice.
Here we consider the pure and the random ferromagnetic Ising model with the classical energy
\begin{eqnarray}
U({ C}) = -\sum_{i<j} J_{ij} S_i S_j
\label{Uspin}
\end{eqnarray}
to study the properties of stochastic dynamics satisfying detailed balance.

\subsection{ Dynamics satisfying detailed balance }

The stochastic dynamics is defined by the master equation 
\begin{eqnarray}
\frac{ dP_t \left({ C} \right) }{dt}
= \sum_{ C '} P_t \left({ C}' \right) 
W \left({ C}' \to  { C}  \right) 
 -  P_t \left({ C} \right) W_{out} \left( { C} \right)
\label{master}
\end{eqnarray}
that describes the time evolution of the
probability $P_t ({ C} ) $ to be in  configuration ${ C}$
 at time t.
The notation $ W \left({ C}' \to  { C}  \right) $ 
represents the transition rate per unit time from configuration 
${ C}'$ to ${ C}$, and 
\begin{eqnarray}
W_{out} \left( { C} \right)  \equiv
 \sum_{ { C} '} W \left({ C} \to  { C}' \right) 
\label{wcout}
\end{eqnarray}
represents the total exit rate out of configuration ${ C}$.

The convergence towards Boltzmann equilibrium at temperature $T=\frac{1}{\beta}$
in any finite system
\begin{eqnarray}
P_{eq}({ C}) = \frac{ e^{- \beta U({ C})} }{Z}
\end{eqnarray}
where $Z$ is the partition function
\begin{eqnarray}
Z = \sum_{ C}  e^{- \beta U({ C})}
\label{partition}
\end{eqnarray}
can be ensured by imposing the detailed balance property
\begin{eqnarray}
e^{- \beta U({ C})}   W \left(  C \to  C '  \right)
= e^{- \beta U({ C '})}   W \left(  C' \to  C   \right)
\label{detailed}
\end{eqnarray}
It is thus convenient to parametrize the transition rates as
\begin{eqnarray}
   W \left(  C \to  C '  \right)
= G\left(  C ,  C '  \right) e^{- \frac{\beta}{2} \left[  U({ C '})- U({ C}) \right] }  
\label{gdetailed}
\end{eqnarray}
where 
\begin{eqnarray}
 G\left(  C ,  C '  \right)= G\left(  C' ,  C   \right)
= \sqrt{W \left( { C} \to { C '} \right) W \left( { C '} \to { C} \right) }  
\label{defG}
\end{eqnarray}
is a symmetric positive function of the two configurations. 
Near zero temperature, it is convenient to introduce 
the notion of dynamical barrier ${\cal B}$ defined by the asymptotic behavior
\begin{eqnarray}
G(C,C') \oppropto_{\beta \to +\infty} e^{- \beta {\cal B}(C,C') }
\label{defbarrier}
\end{eqnarray}

\subsection{ Associated quantum Hamiltonian}

The standard similarity transformation (see for instance the textbooks
 \cite{gardiner,vankampen,risken} or the works concerning spin models
 \cite{glauber,felderhof,siggia,kimball,peschel,us_conjugate,castelnovo,us_rgdyn})
\begin{eqnarray}
P_t ( { C} ) \equiv e^{-  \frac{\beta}{2} U( C ) } \psi_t ({ C} )
=   e^{-  \frac{\beta}{2} U( C ) } <{ C} \vert  \psi_t  >
\label{relationPpsi}
\end{eqnarray}
transforms the master equation of Eq. \ref{master}
into the imaginary-time  Schr\"odinger equation
for the ket  $\vert  \psi_t  >$ 
\begin{eqnarray}
\frac{ d }{dt} \vert  \psi_t  > = -H \vert  \psi_t  > 
\label{Hquantum}
\end{eqnarray}
with the quantum Hamiltonian 
\begin{eqnarray}
{ H} =  \sum_{{ C},{ C '}}  G({ C} , { C '}) \left[  
e^{- \frac{\beta}{2} \left[  U({ C '})- U({ C} \right] } \vert { C } > < { C } \vert
 - \vert { C '} > < { C } \vert \right]
\label{tight}
\end{eqnarray}

The groundstate energy is $E_0=0$, and the corresponding
eigenvector corresponding to the Boltzmann equilibrium reads
\begin{eqnarray}
\vert  \psi_0 > = \sum_{ C}  \frac{ e^{- \frac{\beta}{2} U({ C}) }}{\sqrt Z}
\vert {  C}  >
\label{psi0}
\end{eqnarray}

The other energies $E_n>0$ determine the relaxation towards equilibrium.
In particular, the lowest non-vanishing energy $E_1$
determines the largest relaxation time $(1/E_1)$ of the system, i.e. the 'equilibrium time' 
 needed to converge towards equilibrium, 
\begin{eqnarray}
t_{eq} \equiv \frac{1}{E_1}
\label{deftaueq}
\end{eqnarray}

\subsection{ Single-spin flip dynamics of Ising models }

We will focus here on single spin-flip dynamics satisfying
detailed balance of Eq. \ref{detailed},
 where the transition rate corresponding to the flip of a single spin $S_k$ reads
\begin{eqnarray}
W \left( S_k \to -S_k \right)
=G^{ini} \left[ h_k=\sum_{i \ne k} J_{ik} S_i \right]   e^{ - \beta S_k \left[ \sum_{i \ne k} J_{ik} S_i \right] }
\label{WG}
\end{eqnarray}
$G^{ini}[h_k]$ is an arbitrary positive even function of the local field $h_k=\sum_{i \ne k} J_{ik} S_i $.
For instance, the Glauber dynamics corresponds to the choice
\begin{eqnarray}
G^{ini}_{Glauber} [h] = \frac{1}
{ 2 \cosh \left( \beta  h \right) }
\label{glauber}
\end{eqnarray}
 The quantum Hamiltonian of Eq. \ref{tight}
reads in terms of Pauli matrices $(\sigma^x,\sigma^z)$
\cite{glauber,felderhof,siggia,kimball,peschel,us_conjugate,castelnovo,us_rgdyn}
\begin{eqnarray}
{\cal H } && 
=  \sum_{ k } G^{ini} \left[ \sum_{i \ne k} J_{ik} \sigma^z_i \right]
  \left( e^{ - \beta \sigma^z_k \left( \sum_{i \ne k} J_{ik} \sigma^z_i \right) }
-   \sigma^x_k \right)
\label{Hgene}
\end{eqnarray}

For ferromagnetic models near zero-temperature, more precisely when the temperature is much smaller than any
ferromagnetic coupling $J_{ij}$
\begin{eqnarray}
0< T \ll J_{ij}
\label{lowT}
\end{eqnarray}
 the thermal equilibrium is dominated by the two ferromagnetic groundstates where all spins
take the same value, and the largest relaxation time $t_{eq}\simeq 1/E_1$ corresponds to the time needed to go
from one groundstate (where all spins take the value $+1$)
to the opposite groundstate (where all spins take the value $-1$).
The aim of the renormalization procedure introduced in \cite{us_rgdyn}
is to preserve the lowest non-vanishing energy $E_1$ of the quantum Hamiltonian.
We have already explained the application of this renormalization 
to the random ferromagnetic chain \cite{us_rgdyn},
to the random ferromagnetic Cayley tree \cite{us_rgdyntree}, 
and to the Dyson hierarchical Ising model \cite{us_rgdyndyson}. 
In the following, we derive the appropriate renormalization rules
  for the diamond hierarchical lattice described above.

\section{ Dynamics of the pure Ising model}

\label{sec_pure}

\subsection{ Principle of the Renormalization for the dynamics }

To analyze the statics of the ferromagnetic Ising model on the diamond lattice,
one has to renormalize {\it bonds } to obtain the RG rules for renormalized couplings.
However here to analyze the single-spin-flip dynamics, we wish to define {\it renormalized spins }
that represent ferromagnetic clusters of spins flipping together.

We start from the diamond lattice with $n \geq 2$ generations, of length $L_n=2^n$, containing
$B_n=(2K)^n$ bonds. All bonds have the same initial ferromagnetic coupling $J$.
All elementary spins have a dynamics characterized by the function $G^{ini}[h]$
of the local field $h$ (Eq \ref{Hgene}).

After one RG step, we wish to have a diamond graph of generation $(n-1)$,
where there is an alternation of initial spins, with a dynamics still characterized
by the function $G^{ini}[h]$, and of renormalized spins $S_{R_1}$ (representing clusters of $(2K+1)$ initial spins),
whose dynamics is represented by some renormalized amplitude $G_{R1}$.
The renormalized bonds have for renormalized couplings $J_{R_1}=K J$.

More generally, after $p$ RG steps with $1 \leq p \leq n-1$, we wish to have a diamond graph of generation $(n-p)$,
where there is an alternation of initial spins, with a dynamics still characterized
by the function $G^{ini}[h]$, and of renormalized spins $S_{R_p}$
 (representing clusters of $(2K+1)^p$ initial spins),
whose dynamics is represented by some renormalized amplitude $G_{R_p}$.
The renormalized bonds have for renormalized couplings 
\begin{eqnarray}
J_{R_p}=K^p J
\label{JRp}
\end{eqnarray}

In particular, after $p=n-1$ RG steps, we wish to have a diamond graph of generation $1$,
where the two boundary spins are initial spins, with a dynamics still characterized
by the function $G^{ini}[h]$, and where the $K$ internal renormalized spins $S_{R_{n-1}}$
 (representing clusters of $(2K+1)^{n-1}$ initial spins),
whose dynamics is represented by some renormalized amplitude $G_{R_{n-1}}$.
The renormalized bonds have for renormalized couplings $J_{R_{n-1}}=K^{n-1} J$.

The last RG $p=n$ RG step is thus special : the $(K+2)$ remaining spins
are grouped together into a single renormalized spin $S_{R_{n}}$ 
(representing the whole sample of $(K (2K+1)^{n-1}+2)$ initial spins)
whose dynamics is represented by some renormalized amplitude  $G_{R_n}^{last}$.

Let us now explain how the RG rule for the renormalized amplitude $G_R$ describing the dynamics
can be obtained for the special last RG step $p=n$  and for the bulk RG steps $1 \leq p \leq n-1$
respectively.

\subsection{ RG rule for the last RG step  $p=n$ }

\begin{figure}[htbp]
\includegraphics[height=6cm]{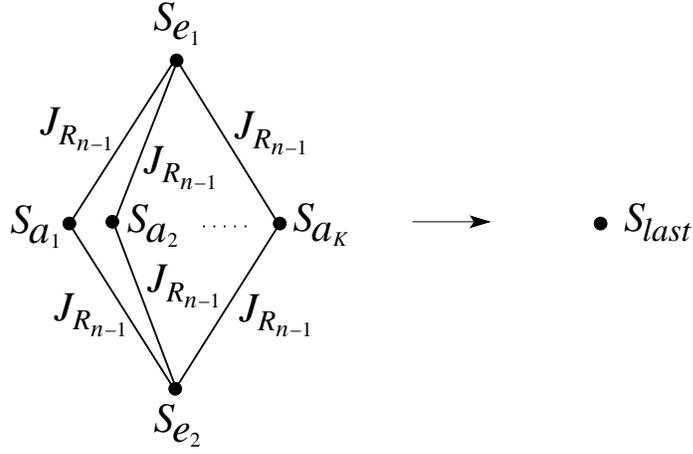}
\hspace{1cm}
\caption{ Last RG step for the dynamics of the pure Ising model : notations for the quantum Hamiltonian of Eq. \ref{HKpurbord}   }
\label{figpurebord}
\end{figure}

The last RG step $p=n$ shown on Fig. \ref{figpurebord}
involves two boundary spins $S_{e_1}$ and $S_{e_2}$ and $K$ internal spins $(S_{a_1},S_{a_2},..S_{a_K})$
with the following classical energy (Eq. \ref{Uspin})
\begin{eqnarray}
U({ C}) = - J_{R_{n-1}} \sum_{i=1}^K  S_{a_i} (  S_{e_1}+  S_{e_2})
\label{Uspinbordpur}
\end{eqnarray}

The quantum Hamiltonian of Eq. \ref{Hgene} associated to the single-spin-flip dynamics reads
\begin{eqnarray}
 H_{K+2} && = G^{ini} \left[ J_{R_{n-1}}\sum_{i=1}^K  \sigma^z_{a_i}  \right]
 \left( e^{ - \beta  J_{R_{n-1}} \sigma^z_{e_1} \sum_{i=1}^K  \sigma^z_{a_i}   } -   \sigma^x_{e_1} \right)
\nonumber \\
&& + G^{ini} \left[J_{R_{n-1}} \sum_{i=1}^K   \sigma^z_{a_i}  \right]
 \left( e^{ - \beta J_{R_{n-1}} \sigma^z_{e_2} \sum_{i=1}^K   \sigma^z_{a_i}   } -   \sigma^x_{e_2} \right)
\nonumber \\
&& +  
 \sum_{i=1}^K G_{R_{n-1}} \left[ J_{R_{n-1}}  (\sigma^z_{e_1}+ \sigma^z_{e_2} ) \right]
 \left( e^{ - \beta J_{R_{n-1}} \sigma^z_{a_i}  (  \sigma^z_{e_1}+ \sigma^z_{e_2} )  } -   \sigma^x_{a_i} \right)
\label{HKpurbord}
\end{eqnarray}

We are interested in the lowest non-vanishing eigenvalue $E_1>0$.
The eigenvalue equation 
\begin{eqnarray}
0 && = (H_{K+2}- E_1) \vert  \vert \psi_1 > 
\label{eigenpsi1lastpurdef}
\end{eqnarray}
for the corresponding eigenvector in the basis
\begin{eqnarray}
\vert \psi_1 > = \sum_{S_{e_1}=\pm} \sum_{S_{e_2}=\pm}  \sum_{S_{a_1}=\pm} .. \sum_{S_{a_K}=\pm} 
\psi_1( S_{e_1},S_{e_2}, S_{a_1} .. S_{a_K} ) 
\vert S_{e_1},S_{e_2}, S_{a_1} .. S_{a_K}>
\label{psi1lastpur}
\end{eqnarray}
reads
\begin{eqnarray}
0 && = [ G^{ini} \left[ J_{R_{n-1}}\sum_{i=1}^K  S_{a_i}  \right]
  e^{ - \beta  J_{R_{n-1}} S_{e_1} \sum_{i=1}^K  S_{a_i}   }
\nonumber \\
&& + G^{ini} \left[J_{R_{n-1}} \sum_{i=1}^K   S_{a_i}  \right]
 e^{ - \beta J_{R_{n-1}} S_{e_2} \sum_{i=1}^K   S_{a_i}   }
\nonumber \\
&& +  
 \sum_{i=1}^K G_{R_{n-1}} \left[ J_{R_{n-1}}  (S_{e_1}+ S_{e_2} ) \right]
  e^{ - \beta J_{R_{n-1}} S_{a_i}  (  S_{e_1}+ S_{e_2} )  } - E_1 ]
\psi_1( S_{e_1},S_{e_2}, S_{a_1} .. S_{a_K} ) 
\nonumber \\
&& - 
G^{ini} \left[ J_{R_{n-1}}\sum_{i=1}^K  S_{a_i}  \right]
    \psi_1( - S_{e_1},S_{e_2}, S_{a_1} .. S_{a_K} ) 
\nonumber \\
&& - G^{ini} \left[J_{R_{n-1}} \sum_{i=1}^K   S_{a_i}  \right]
    \psi_1( S_{e_1},- S_{e_2}, S_{a_1} .. S_{a_K} ) 
\nonumber \\
&& -  
 \sum_{i=1}^K G_{R_{n-1}} \left[ J_{R_{n-1}}  (S_{e_1}+ S_{e_2} ) \right]
     \psi_1( S_{e_1},S_{e_2}, S_{a_1} ,..., -S_{a_i},.. S_{a_K} ) 
\label{eigenpsi1lastpur}
\end{eqnarray}

Let us now use the symmetry between the $K$ spins $S_{a_i}$
to note $\phi(S_{e_1},k,S_{e_2})$ the components $\psi_1( S_{e_1},S_{e_2}, S_{a_1} .. S_{a_K} ) $ 
where $k \in {0,1,..,K}$ spins among $(S_{a_1},..,S_{a_K} )$ take the value $(-)$
\begin{eqnarray}
\psi_1( S_{e_1},S_{e_2}, S_{a_1} .. S_{a_K} )=\phi_{1}(S_{e_1}, k=\sum_{i=1}^K \frac{1-S_{a_i}}{2}, S_{e_2})
\label{rksym}
\end{eqnarray}
Eq \ref{eigenpsi1lastpur} becomes
\begin{eqnarray}
&& 0  = [ G^{ini} \left[ J_{R_{n-1}} (K-2k) \right]
  e^{ - \beta  J_{R_{n-1}} S_{e_1} (K-2k)    }
 + G^{ini} \left[J_{R_{n-1}} (K-2k)   \right]
 e^{ - \beta J_{R_{n-1}} S_{e_2} (K-2k)    }
\nonumber \\
&& +  
 k G_{R_{n-1}} \left[ J_{R_{n-1}}  (S_{e_1}+ S_{e_2} ) \right]
  e^{  \beta J_{R_{n-1}}   (  S_{e_1}+ S_{e_2} )  } 
+\nonumber \\
&& (K-k) G_{R_{n-1}} \left[ J_{R_{n-1}}  (S_{e_1}+ S_{e_2} ) \right]
  e^{ - \beta J_{R_{n-1}}   (  S_{e_1}+ S_{e_2} )  } - E_1 ]
\phi_1( S_{e_1},k, S_{e_2}) 
\nonumber \\
&& - 
G^{ini} \left[ J_{R_{n-1}} (K-2k)    \right]
    \phi_1( - S_{e_1},k, S_{e_2} ) 
\nonumber \\
&& - G^{ini} \left[J_{R_{n-1}} (K-2k)   \right]
    \phi_1( S_{e_1},k, - S_{e_2} ) 
\nonumber \\
&& -  
 k G_{R_{n-1}} \left[ J_{R_{n-1}}  (S_{e_1}+ S_{e_2} ) \right]
     \phi_1( S_{e_1},k-1,S_{e_2} ) 
\nonumber \\
&& -  
(K-k) G_{R_{n-1}} \left[ J_{R_{n-1}}  (S_{e_1}+ S_{e_2} ) \right]
     \phi_1( S_{e_1},k+1,S_{e_2} ) 
\label{eigenpsi1lastpursym}
\end{eqnarray}

For $E_0=0$, the (non-normalized) groundstate is known to be given by Eq. \ref{psi0}
with the classical energy of Eq. \ref{Uspinbordpur}
\begin{eqnarray}
\phi_0( S_{e_1},k, S_{e_2} ) = e^{ \frac{\beta}{2} J_{R_{n-1}} (K-2k) (  S_{e_1}+  S_{e_2}) }
\label{psizerobordpur}
\end{eqnarray}
so it is convenient to look for the solution of Eq. \ref{eigenpsi1lastpur} via the amplitude
\begin{eqnarray}
A ( S_{e_1},k, S_{e_2} ) = \frac{\phi_1( S_{e_1},k, S_{e_2} )}{\phi_0( S_{e_1},k, S_{e_2})}
\label{amplibordpur}
\end{eqnarray}
that satisfies
\begin{eqnarray}
&& 0  = [  G^{ini} \left[ J_{R_{n-1}} (K-2k) \right]
  e^{ - \beta  J_{R_{n-1}}(K-2k)  S_{e_1}     }
 + G^{ini} \left[J_{R_{n-1}} (K-2k)   \right]
 e^{ - \beta J_{R_{n-1}}(K-2k) S_{e_2}     }
\nonumber \\
&& +  
 k G_{R_{n-1}} \left[ J_{R_{n-1}}  (S_{e_1}+ S_{e_2} ) \right]
  e^{  \beta J_{R_{n-1}}   (  S_{e_1}+ S_{e_2} )  } 
\nonumber \\
&& + (K-k) G_{R_{n-1}} \left[ J_{R_{n-1}}  (S_{e_1}+ S_{e_2} ) \right]
  e^{ - \beta J_{R_{n-1}}   (  S_{e_1}+ S_{e_2} )  } 
- E_1  ] A( S_{e_1},k, S_{e_2}) 
\nonumber \\
&& - 
G^{ini} \left[ J_{R_{n-1}} (K-2k)    \right]
 e^{ - \beta J_{R_{n-1}} (K-2k)  S_{e_1} }   A( - S_{e_1},k, S_{e_2} ) 
\nonumber \\
&& - G^{ini} \left[J_{R_{n-1}} (K-2k)   \right]
 e^{ - \beta J_{R_{n-1}}(K-2k)   S_{e_2} }    A( S_{e_1},k, - S_{e_2} ) 
\nonumber \\
&& -  
 k G_{R_{n-1}} \left[ J_{R_{n-1}}  (S_{e_1}+ S_{e_2} ) \right]
 e^{ \beta J_{R_{n-1}}  (  S_{e_1}+  S_{e_2}) }      A( S_{e_1},k-1,S_{e_2} )
\nonumber \\
&& -  
(K-k) G_{R_{n-1}} \left[ J_{R_{n-1}}  (S_{e_1}+ S_{e_2} ) \right]
  e^{ - \beta J_{R_{n-1}} (  S_{e_1}+  S_{e_2}) }     A( S_{e_1},k+1,S_{e_2} )
\label{eigenpsi1lastpursymA}
\end{eqnarray}

Let us now consider the dynamical path associated to the swap of a domain-wall between the two ferromagnetic
groundstates with the following notations for configurations
\begin{eqnarray}
C_0 &&  =\{S_{e_1}=+1,0,S_{e_2}=+1 \}
\nonumber \\
C_1 &&  =\{S_{e_1}=-1,0,S_{e_2}=+1 \}
\nonumber \\
C_2 &&  =\{S_{e_1}=-1,1,S_{e_2}=+1 \}
\nonumber \\
C_p &&  =\{S_{e_1}=-1,p-1,S_{e_2}=+1 \}
\nonumber \\
C_K &&  =\{S_{e_1}=-1,K-1,S_{e_2}=+1 \}
\nonumber \\
C_{K+1} &&  =\{S_{e_1}=-1,K,S_{e_2}=+1 \}
\nonumber \\
C_{K+2} &&  =\{S_{e_1}=-1,K,S_{e_2}=-1 \}
\label{configbordpur}
\end{eqnarray}
The physical meaning is that the transition from $C_0$ to $C_1$ corresponds to the entrance of a domain-wall at the boundary $e_1$, the transitions between the configurations $(C_1,C_2,..,C_{K+1})$ correspond to the displacement of this domain-wall, and finally the transition from $C_{K+1}$ to $C_{K+2}$ corresponds to the exit of the domain-wall at the boundary $e_2$.

The corresponding amplitudes $A(C_q)$ for this dynamical path satisfy (Eq \ref{eigenpsi1lastpursymA})
for $2 \leq q \leq K$
\begin{eqnarray}
 \nonumber 0  = && \left( 
 (q-1) G_{R_{n-1}} \left[ 0 \right]
+ (K+1-q) G_{R_{n-1}} \left[ 0 \right]
- E_1 \right) A( C_{q}) 
 \nonumber \\ &&
 -  
 (q-1) G_{R_{n-1}} \left[ 0 \right]     A( C_{q-1} )
-  
(K+1-q) G_{R_{n-1}} \left[ 0  \right]   A( C_{q+1} )
\label{eigenpsi1lastpursymADW}
\end{eqnarray}
and for $q=0$, $q=1$, $q=K+1$ and $q=K+2$
\begin{eqnarray}
 0  = && [ G^{ini} \left[ K J_{R_{n-1}}  \right]
  e^{ - \beta K  J_{R_{n-1}}      } 
- E_1 ] A( C_0)  - 
G^{ini} \left[ K J_{R_{n-1}}    \right]
 e^{ - \beta K J_{R_{n-1}}   }   A( C_1 ) 
\nonumber \\
 0  = &&[ G^{ini} \left[ K J_{R_{n-1}} \right]
  e^{  \beta K  J_{R_{n-1}}      }
+ K G_{R_{n-1}} \left[ 0 \right]
- E_1 ] A( C_1 ) 
 \nonumber \\ &&  - 
G^{ini} \left[ K J_{R_{n-1}}     \right]
 e^{  \beta K J_{R_{n-1}}   }   A( C_0 ) 
 -   K  G_{R_{n-1}} \left[ 0 \right]
      A( C_2 )
\nonumber \\
 0  = &&[ 
  G^{ini} \left[ K J_{R_{n-1}}  \right]
 e^{  \beta K J_{R_{n-1}}      }
 +  
 K G_{R_{n-1}} \left[ 0\right]
- E_1 ] A( C_{K+1})  
 \nonumber \\ && - G^{ini} \left[K J_{R_{n-1}}   \right]
 e^{  \beta K J_{R_{n-1}}   }    A( C_{K+2} ) 
 -  
 K G_{R_{n-1}} \left[ 0 \right]
     A( C_K )
\nonumber \\
 0  = && [  G^{ini} \left[ K J_{R_{n-1}}   \right]
 e^{ -  \beta K J_{R_{n-1}}      }
- E_1 ] A( C_{K+2} ) 
 - G^{ini} \left[ K J_{R_{n-1}}    \right]
 e^{ - K \beta J_{R_{n-1}}  }    A( C_{K+1} ) 
\label{eigenpsi1lastpursymADWbis}
\end{eqnarray}

For this effective one-dimensional problem with the effective transition rates
\begin{eqnarray}
W^{eff}(C_0 \to C_{1}) && = G^{ini} \left[ K J_{R_{n-1}}    \right]  e^{ - \beta K J_{R_{n-1}}   }
\nonumber \\
W^{eff}(C_q \to C_{q+1}) && = (K+1-q) G_{R_{n-1}} \left[ 0  \right]  \ \ \ \ {\rm for } \ \ 1 \leq q \leq K
\nonumber \\
W^{eff}(C_{K+1} \to C_{K+2}) && = G^{ini} \left[K J_{R_{n-1}}   \right] e^{  \beta K J_{R_{n-1}}   }
\label{effwbord}
\end{eqnarray}
and
\begin{eqnarray}
W^{eff}(C_{1} \to C_{0}) && = G^{ini} \left[ K J_{R_{n-1}}     \right] e^{  \beta K J_{R_{n-1}}   } 
\nonumber \\
W^{eff}(C_q \to C_{q-1}) && =  (q-1) G_{R_{n-1}} \left[ 0 \right]  \ \ \ \  {\rm for } \ \ 2 \leq q \leq K+1 
\nonumber \\
W^{eff}(C_{K+2} \to C_{K+1}) && =  G^{ini} \left[ K J_{R_{n-1}}    \right] e^{ - K \beta J_{R_{n-1}}  } 
\label{effwbordbis}
\end{eqnarray}
we may use Eq. \ref{Grfinalw} of the Appendix to obtain the renormalized amplitude $G_{R_nlast}^{(e_1,e_2)}=
G_R ({ C_0} , { C_{K+2}}) $ when the domain-wall enters by the boundary $e_1$ and exits by the boundary $e_2$
\begin{eqnarray}
 \frac{1}{G_{R_nlast}^{(e_1,e_2)}} 
&& =  \frac{ e^{ \frac{\beta}{2} \left[U(C_{0})- U(C_{K+2} )\right] } }
{ W^{eff}(C_0 \to C_1) }
\left[  1+\sum_{m=1}^{K+1} \prod_{q=1}^{m} \frac{W^{eff}(C_q \to C_{q-1})}{W^{eff}(C_q \to C_{q+1})} \right]
\nonumber \\
&& =  \frac{ 2e^{  \beta K J_{R_{n-1}}   } }
{  G^{ini} \left[ K J_{R_{n-1}}    \right]  }
+  \frac{ e^{ 2 \beta K J_{R_{n-1}}   } }{ G_{R_{n-1}}\left[ 0 \right]} \sum_{m=1}^{K}  \frac{ (m-1)! (K-m)! }
 { K! }  
\label{Grfinalwbord}
\end{eqnarray}
Taking into account the other case where the domain-wall enters by the boundary $e_2$ and exits by the boundary $e_1$, which actually gives the same contribution
\begin{eqnarray}
G_{R_nlast}^{(e_1,e_2)}=G_{R_nlast}^{(e_2,e_1)}
\label{GlastK1purbis}
\end{eqnarray}
we obtain that the final total amplitude $G_{R_n}^{last}= G_{R_nlast}^{(e_1,e_2)}+G_{R_nlast}^{(e_2,e_1)}=2 G_{R_nlast}^{(e_1,e_2)} $ reads
\begin{eqnarray}
 \frac{1}{G_{R_n}^{last}} 
 =  \frac{ e^{  \beta K J_{R_{n-1}}   } }
{  G^{ini} \left[ K J_{R_{n-1}}    \right]  }
+  \frac{ e^{ 2 \beta K J_{R_{n-1}}   } }{ 2 G_{R_{n-1}}\left[ 0 \right]} \sum_{k=0}^{K-1}  \frac{ k! (K-1-k)! }
 { K! }  
\label{Glastpurbordfinal}
\end{eqnarray}

\subsection{ RG rule for the bulk RG steps  $1 \leq p \leq n-1$ }

\begin{figure}[htbp]
\includegraphics[height=8cm]{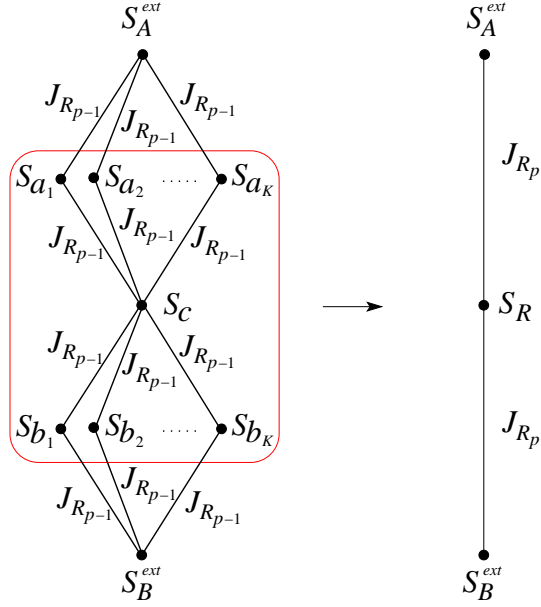}
\hspace{1cm}
\caption{ Bulk RG step for the dynamics of the pure Ising model : notations for the quantum Hamiltonian of Eq. \ref{HKbulkpur}   }
\label{figpurebulk}
\end{figure}

The bulk RG step $p$ shown on Fig. \ref{figpurebulk}
involves two external spins $S_A^{ext} $ and $S_B^{ext} $,
and $(2K+1)$ internal spins $(S_{a_1},S_{a_2},.,S_{a_K},S_c,S_{b_1},S_{b_2},.,S_{b_K})$
with the following classical energy (Eq. \ref{Uspin})
\begin{eqnarray}
U({ C}) = - J_{R_{p-1}} \sum_{i=1}^K  S_{a_i} (  S_c+  S_A^{ext}) - J_{R_{p-1}} \sum_{i=1}^K  S_{b_i} (  S_c+ S_B^{ext})
\label{Uspinbulkpur}
\end{eqnarray}

The quantum Hamiltonian of Eq. \ref{Hgene} associated to the single-spin-flip dynamics reads
\begin{eqnarray}
 H_{2K+1} &&  = G^{ini} \left[ J_{R_{p-1}} \sum_{i=1}^K (  \sigma^z_{a_i}+ \sigma^z_{b_i} ) \right]
 \left( e^{ - \beta J_{R_{p-1}}  \sigma^z_c \sum_{i=1}^K (  \sigma^z_{a_i}+\sigma^z_{b_i} )  } -   \sigma^x_c \right)
\nonumber \\
&& +  
 \sum_{i=1}^K G_{R_{p-1}}\left[ J_{R_{p-1}} ( \sigma^z_c+ S_A^{ext} ) \right]
 \left( e^{ - \beta J_{R_{p-1}} \sigma^z_{a_i}  (  \sigma^z_c+ S_A^{ext} )  } -   \sigma^x_{a_i} \right)
\nonumber \\
&& +  
 \sum_{i=1}^K  G_{R_{p-1}}\left[  J_{R_{p-1}} ( \sigma^z_c+ S_B^{ext}) \right]
\left( e^{ - \beta J_{R_{p-1}} \sigma^z_{b_i}  ( \sigma^z_c+ S_B^{ext} )  } -   \sigma^x_{b_i} \right)
\label{HKbulkpur}
\end{eqnarray}

Let us now focus on the external Domain-Wall conditions 
\begin{eqnarray}
S_A^{ext} && =-1
\nonumber \\
S_B^{ext} && =+1
\label{dwextab}
\end{eqnarray}
and take into account the symmetry between the $K$ spins $S_{a_i}$,
and the symmetry between the $K$ spins $S_{b_i}$
to note $\phi(k_a,k_b)$ the components of $\psi_1$
where $k_a \in {0,1,..,K}$ spins among $(S_{a_1},..,S_{a_K} )$ take the value $(-)$,
and where  $k_a \in {0,1,..,K}$ spins among $(S_{a_1},..,S_{a_K} )$ take the value $(-)$
\begin{eqnarray}
\psi_1(S_A^{ext} =-1,  S_{a_1}, .. ,S_{a_K},S_c,S_{b_1},..,S_{b_K},S_B^{ext} && =+1 )=
\phi_{1}( k_a=\sum_{i=1}^K \frac{1-S_{a_i}}{2}, S_{c},k_b=\sum_{i=1}^K \frac{1-S_{b_i}}{2} )
\label{rksymab}
\end{eqnarray}

Then the eigenvalue equation $0=(H_{2K+1}-E_1) \vert \psi_1>$ reads
\begin{eqnarray}
0 &&  = (G^{ini} \left[ J_{R_{p-1}}  (  2K-2k_a-2 k_b ) \right]
  e^{ - \beta J_{R_{p-1}} (  2K-2k_a-2 k_b ) S_c   } 
\nonumber \\
&& +  
 k_a G_{R_{p-1}}\left[ J_{R_{p-1}} ( S_c-1 ) \right]
  e^{  \beta J_{R_{p-1}}   (  S_c-1 )  } 
\nonumber \\
&& +  
 (K-k_a) G_{R_{p-1}}\left[ J_{R_{p-1}} ( S_c-1 ) \right]
  e^{ - \beta J_{R_{p-1}}   (  S_c-1 )  } 
\nonumber \\
&& +  
 k_b  G_{R_{p-1}}\left[  J_{R_{p-1}} ( S_c+ 1) \right]
 e^{  \beta J_{R_{p-1}}   ( S_c+ 1 )  }
\nonumber \\
&& +  
 (K-k_b)  G_{R_{p-1}}\left[  J_{R_{p-1}} ( S_c+ 1) \right]
 e^{ - \beta J_{R_{p-1}}   ( S_c+ 1 )  } - E_1 )
\phi_1(k_a,S_c,k_b)
\nonumber \\
&&
- G^{ini} \left[ J_{R_{p-1}} (  2K-2k_a-2 k_b ) \right]
     \phi_1(k_a,-S_c,k_b)
\nonumber \\
&& - 
 k_a G_{R_{p-1}}\left[ J_{R_{p-1}} ( S_c-1 ) \right]
     \phi_1(k_a-1,S_c,k_b)
\nonumber \\
&& - 
 (K-k_a)  G_{R_{p-1}}\left[ J_{R_{p-1}} ( S_c-1 ) \right]
     \phi_1(k_a+1,S_c,k_b)
\nonumber \\
&& -  
 k_b  G_{R_{p-1}}\left[  J_{R_{p-1}} ( S_c+ 1) \right]
   \phi_1(k_a,S_c,k_b-1)
\nonumber \\
&& -  
 (K-k_b) G_{R_{p-1}}\left[  J_{R_{p-1}} ( S_c+ 1) \right]
   \phi_1(k_a,S_c,k_b+1)
\label{eigenbulkpur}
\end{eqnarray}

For $E_0=0$, the (non-normalized) groundstate is known to given by Eq. \ref{psi0}
with the classical energy of Eq. \ref{Uspinbulkpur}
\begin{eqnarray}
\phi_0( k_a,S_c,k_b ) = e^{ \frac{\beta}{2} J_{R_{p-1}} \left[ S_c (2 K-2 k_a-2k_b) + (2 k_a-2 k_b) \right]}
\label{psizeropur}
\end{eqnarray}
so it is convenient to look for the solution of Eq. \ref{eigenpsi1lastpur} via the amplitude
\begin{eqnarray}
A (k_a,S_c,k_b  ) = \frac{\phi_1( k_a,S_c,k_b )}{\phi_0( k_a,S_c,k_b )}
\label{amplipur}
\end{eqnarray}
that satisfies
\begin{eqnarray}
0 &&  = (G^{ini} \left[ J_{R_{p-1}}  (  2K-2k_a-2 k_b ) \right]
  e^{ - \beta J_{R_{p-1}} (  2K-2k_a-2 k_b ) S_c   } 
\nonumber \\
&& +  
 k_a G_{R_{p-1}}\left[ J_{R_{p-1}} ( S_c-1 ) \right]
  e^{  \beta J_{R_{p-1}}   (  S_c-1 )  } 
+  
 (K-k_a) G_{R_{p-1}}\left[ J_{R_{p-1}} ( S_c-1 ) \right]
  e^{ - \beta J_{R_{p-1}}   (  S_c-1 )  } 
\nonumber \\
&& +  
 k_b  G_{R_{p-1}}\left[  J_{R_{p-1}} ( S_c+ 1) \right]
 e^{  \beta J_{R_{p-1}}   ( S_c+ 1 )  }
+  
 (K-k_b)  G_{R_{p-1}}\left[  J_{R_{p-1}} ( S_c+ 1) \right]
 e^{ - \beta J_{R_{p-1}}   ( S_c+ 1 )  } - E_1 )
 A(k_a,S_c,k_b)
\nonumber \\
&&
- G^{ini} \left[ J_{R_{p-1}} (  2K-2k_a-2 k_b ) \right]
 e^{- \beta J_{R_{p-1}} (2 K-2 k_a-2k_b) S_c }
      A (k_a,-S_c,k_b)
\nonumber \\
&& - 
 k_a G_{R_{p-1}}\left[ J_{R_{p-1}} ( S_c-1 ) \right]
 e^{ \beta J_{R_{p-1}}  (S_c-1) }     A(k_a-1,S_c,k_b)
 \nonumber \\ && 
- 
 (K-k_a)  G_{R_{p-1}}\left[ J_{R_{p-1}} ( S_c-1 ) \right]
   e^{ - \beta J_{R_{p-1}}  (S_c-1) }    A (k_a+1,S_c,k_b)
\nonumber \\
&& -  
 k_b  G_{R_{p-1}}\left[  J_{R_{p-1}} ( S_c+ 1) \right]
 e^{ \beta J_{R_{p-1}}  (S_c+1) }    A(k_a,S_c,k_b-1)
 \nonumber \\ &&
 -  
 (K-k_b) G_{R_{p-1}}\left[  J_{R_{p-1}} ( S_c+ 1) \right]
  e^{ - \beta J_{R_{p-1}}  (S_c+1) }    A(k_a,S_c,k_b+1)
\label{eigenpsi1pursymA}
\end{eqnarray}

Let us consider the dynamical path along the following $(2K+2)$ configurations
\begin{eqnarray}
C_0 &&  =\{ 0,S_{c}=+1,0 \}
\nonumber \\
C_1 &&  =\{ 1,S_{c}=+1,0 \}
\nonumber \\
C_2 &&  =\{ 2,S_{c}=+1,0 \}
\nonumber \\
C_K &&  =\{ K,S_{c}=+1,0 \}
\nonumber \\
C_{K+1} &&  =\{ K,S_{c}=-1,0 \}
\nonumber \\
C_{K+2} &&  =\{ K,S_{c}=-1,1 \}
\nonumber \\
C_{K+3} &&  =\{ K,S_{c}=-1,2 \}
\nonumber \\
C_{2K} &&  =\{ K,S_{c}=-1,K-1 \}
\nonumber \\
C_{2K+1} &&  =\{ K,S_{c}=-1,K \}
\label{configbulkpur}
\end{eqnarray}
in order to describe the motion of a domain-wall corresponding to the boundary conditions of Eq. \ref{dwextab}.

Then Eq. \ref{eigenbulkpur} becomes for $0 \leq q \leq K-1$
\begin{eqnarray}
0 &&  = (
 q G_{R_{p-1}}\left[ 0 \right]
+  
 (K-q) G_{R_{p-1}}\left[0 \right]
 - E_1 )
 A(C_q) - 
 q G_{R_{p-1}}\left[0 \right]    A(C_{q-1})
- 
 (K-q)  G_{R_{p-1}}\left[ 0 \right]    A (C_{q+1})
\label{eigenbulkpurkaq}
\end{eqnarray}
for $q=K$
\begin{eqnarray}
0 &&  = (G^{ini} \left[ 0 \right]
  +  
 K G_{R_{p-1}}\left[ 0 \right]
 - E_1 )
 A(C_K)
- G^{ini} \left[ 0 \right]
      A (C_{K+1})
 - 
 K G_{R_{p-1}}\left[ 0 \right]    A(C_{K-1})
\label{eigenbulkpurc}
\end{eqnarray}
for $q=K+1 $
\begin{eqnarray}
0 &&  = (G^{ini} \left[ 0 \right]
+
 K  G_{R_{p-1}}\left[ 0 \right]
  - E_1 )
 A(C_{K+1})
- G^{ini} \left[0 \right]
      A (C_K)
 -  
K G_{R_{p-1}}\left[ 0 \right]   A(C_{K+2})
\label{eigenpsi1pursymAcbis}
\end{eqnarray}
and for $q=K+1+k$ for $1 \leq k \leq K$
\begin{eqnarray}
0 &&  = 
 k  G_{R_{p-1}}\left[  0 \right]
+  
 (K-k)  G_{R_{p-1}}\left[ 0 \right] - E_1 )
 A(C_{K+1+k})
 -  
 k  G_{R_{p-1}}\left[ 0 \right]   A(C_{K+k})
 -  
 (K-k) G_{R_{p-1}}\left[ 0 \right]    A(C_{K+2+k})
\label{eigenpsi1pursymAqprime}
\end{eqnarray}

For this effective one-dimensional problem with the effective transition rates
\begin{eqnarray}
W^{eff}(C_q \to C_{q+1}) && = (K-q)  G_{R_{p-1}}\left[ 0 \right]  \ \ \ \ {\rm for } \ \ 0 \leq q \leq K-1
\nonumber \\
W^{eff}(C_K \to C_{K+1}) && = G^{ini} \left[ 0 \right]
\nonumber \\
W^{eff}(C_{q} \to C_{q+1}) && = (2K+1-q) G_{R_{p-1}}\left[ 0 \right]  \ \ \ \ {\rm for } \ \ K+1 \leq q \leq 2K
\label{effwbulk}
\end{eqnarray}
and
\begin{eqnarray}
W^{eff}(C_q \to C_{q-1}) && = q  G_{R_{p-1}}\left[ 0 \right]  \ \ \ \  {\rm for } \ \ 1 \leq q \leq K
\nonumber \\
W^{eff}(C_{K+1} \to C_{K}) && = G^{ini} \left[ 0 \right]
\nonumber \\
W^{eff}(C_{q} \to C_{q-1}) && = (q-K-1) G_{R_{p-1}}\left[ 0 \right]  \ \ \ \  {\rm for } \ \ K+2 \leq q \leq 2K+1
\label{effwbulkbis}
\end{eqnarray}
we may use Eq. \ref{Grfinalw} of the Appendix to obtain the renormalized amplitude
$G_{R_{p}}\left[ 0 \right]=G_R ({ C_0} , { C_{2K+1}}) $ 
\begin{eqnarray}
\frac{1}{G_{R_{p}}\left[ 0 \right] } 
&& =  \frac{ e^{ \frac{\beta}{2} \left[U(C_{0})- U(C_{2K+1} )\right] } }
{ W^{eff}(C_0 \to C_1) }
\left[  1+\sum_{m=1}^{2K} \prod_{q=1}^{m} \frac{W^{eff}(C_q \to C_{q-1})}{W^{eff}(C_q \to C_{q+1})} \right]
\nonumber \\
&& =    \frac{ 1}{G^{ini} \left[ 0 \right] }
+ \frac{2}{K G_{R_{p-1}}\left[ 0 \right]} \sum_{m=0}^{K-1}  \frac{ (m)! (K-1-m)! }{ (K-1)!  } 
\label{Grfinalwbulk}
\end{eqnarray}

\subsection{ Conclusion   }

Using the combinatorial formula \cite{suminvcombi} concerning the sum of the inverse of binomial coefficients
\begin{eqnarray}
\sum_{m=0}^{K-1}  \frac{ (m)! (K-1-m)! }{ (K-1)!  } \equiv \sum_{m=0}^{K-1}  \frac{ 1 }{ C^m_{K-1}  } 
= \frac{K}{2^K} \sum_{k=1}^K \frac{2^k}{k}
\label{combi}
\end{eqnarray}
and the form of renormalized couplings of Eq. \ref{JRp}
\begin{eqnarray}
J_{R_{n-1}}=K^{n-1} J
\label{JRpn}
\end{eqnarray}
we may rewrite the last RG step of Eq. \ref{Glastpurbordfinal} as
\begin{eqnarray}
 \frac{1}{G_{R_n}^{last}} 
 =  \frac{ e^{  \beta K^n J    } }
{  G^{ini} \left[ K^n J   \right]  }
+  \frac{ e^{ 2 \beta K^n J   } }{ 2^{K+1}  G_{R_{n-1}}\left[ 0 \right]}  \sum_{k=1}^K \frac{2^k}{k}
\label{Glastpurlast}
\end{eqnarray}
and the bulk RG steps  $1 \leq p \leq n-1$ of Eq. \ref{Grfinalwbulk} as 
\begin{eqnarray}
\frac{1}{G_{R_{p}}\left[ 0 \right] } 
 =    \frac{ 1}{G^{ini} \left[ 0 \right] }
+ \frac{1}{ 2^{K-1} G_{R_{p-1}}\left[ 0 \right]}  \sum_{k=1}^K \frac{2^k}{k}
\label{Gpurbulk}
\end{eqnarray}
with the initial condition
\begin{eqnarray}
G_{R_{0}}\left[ 0 \right]=G^{ini} \left[ 0 \right]
\label{inip0}
\end{eqnarray}
In terms of the numerical constant
\begin{eqnarray}
c_K \equiv \frac{1}{ 2^{K-1} }  \sum_{k=1}^K \frac{2^k}{k}
\label{ctek}
\end{eqnarray}
the solution of the recurrence of Eq. \ref{Gpurbulk} reads for  $0 \leq p \leq n-1$
\begin{eqnarray}
\frac{1}{G_{R_{p}}\left[ 0 \right] } 
 =  \frac{ 1}{G^{ini} \left[ 0 \right] } \sum_{q=0}^p (c_K)^q  
\label{Gpurbulksolu}
\end{eqnarray}
so that the final renormalized amplitude of Eq. \ref{Glastpurlast} for the whole sample reads
\begin{eqnarray}
 \frac{1}{G_{R_n}^{last}} 
 =  \frac{ e^{  \beta K^n J    } }
{  G^{ini} \left[ K^n J    \right]  }
+  \frac{ e^{ 2 \beta K^n J   } }{ 4  G^{ini} \left[ 0 \right] }  
 \sum_{q=1}^{n} (c_K)^q  
\label{Glastpursolu}
\end{eqnarray}

The final conclusion is thus that the equilibrium time $t_{eq}^{(n)}$
needed to go 
from one groundstate (where all spins take the value $+1$)
to the opposite groundstate (where all spins take the value $-1$)
reads for the Ising model on the diamond lattice of branching ratio $K$ with $n$ generations reads
\begin{eqnarray}
t_{eq}^{(n)} \simeq \frac{1}{G_{R_n}^{last}} 
 =  \frac{ e^{  \beta K^n J    } }
{  G^{ini} \left[ K^n J   \right]  }
+  \frac{ e^{ 2 \beta K^n J   } }{ 4  G^{ini} \left[ 0 \right] }  
 \sum_{q=1}^{n} (c_K)^q  
\label{teqpursolu}
\end{eqnarray}
In particular, the dynamical barrier ${\cal B}^{(n)} $ defined by the low-temperature exponential behavior reads
in terms of the length $L_n^{2^n}$ and of the fractal dimension $d_f$ (Eqs \ref{lndiamond} and \ref{deffdiamond} )
\begin{eqnarray}
{\cal B}^{(n)} \equiv \lim_{\beta \to +\infty} \frac{t_{eq}^{(n)}}{\beta}  =
 2 J  K^n =   2 J L_n^{d_f-1}
\label{Beqpursolu}
\end{eqnarray}
in agreement with the expected scaling $L^{d_s}$ of the energy cost of an interface of
dimension $d_s=d-1$ in a space of dimension $d$.

Here, we have in addition computed explicitly the prefactors in Eq. \ref{teqpursolu} :
the leading behavior is a power law of the length $L_n$
\begin{eqnarray}
t_{eq}^{(n)} && \simeq   e^{  \beta  2 J L_n^{d_f-1}   } (c_K)^n =  e^{  \beta  2 J L_n^{d_f-1}   } L_n^{\alpha_K}
\label{teqpursolupref}
\end{eqnarray}
where the exponent reads in terms of the constant $c_K$ of Eq. \ref{ctek}
\begin{eqnarray}
 \alpha_K = \frac{ \ln c_K }{ \ln 2} = \frac{\ln \left[\frac{1}{ 2^{K-1} }  \sum_{k=1}^K \frac{2^k}{k} \right]  }{ \ln 2} 
\label{alphaK}
\end{eqnarray}
In particular, for the first values of $K$, we obtain
\begin{eqnarray}
\alpha_{K=1} && = 1
\nonumber \\
\alpha_{K=2} && = 1
\nonumber \\
\alpha_{K=3} && = \frac{ \ln \frac{5}{ 3 } }{\ln 2}
\nonumber \\
\alpha_{K=4} && = \frac{ \ln \frac{4}{ 3 }  }{\ln 2} 
\label{alphaKfirst}
\end{eqnarray}
For $K=1$ corresponding to the one-dimensional chain, the result $\alpha_{K=1} =1$ is in agreement with
previous studies (see \cite{us_rgdyn} and references therein).

\section{ Dynamics of the random ferromagnetic model}

\label{sec_random}

\subsection{ RG rule for the last RG step $p=n$  }

\begin{figure}[htbp]
\includegraphics[height=6cm]{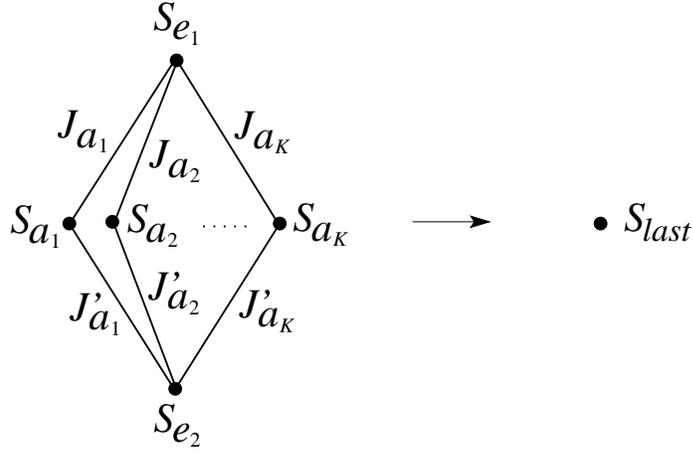}
\hspace{1cm}
\caption{ Last RG step for the dynamics of the disordered Ising model : notations for the quantum Hamiltonian of Eq. \ref{HKdesbord}   }
\label{figdesbord}
\end{figure}

The last RG step $p=n$ shown on Fig. \ref{figdesbord}
involves two boundary spins $S_{e_1}$ and $S_{e_2}$ and $K$ internal spins $(S_{a_1},S_{a_2},..S_{a_K})$
with the following classical energy (Eq. \ref{Uspin})
\begin{eqnarray}
U({ C}) = -\sum_{i=1}^K  S_{a_i} ( J_{a_i} S_{e_1}+ J_{a_i}' S_{e_2})
\label{Uspinbord}
\end{eqnarray}

The quantum Hamiltonian of Eq. \ref{Hgene} associated to the single-spin-flip dynamics reads
\begin{eqnarray}
 H_{K+2} && = G^{ini} \left[ \sum_{i=1}^K  J_{a_i} \sigma^z_{a_i}  \right]
 \left( e^{ - \beta  \sigma^z_{e_1} \sum_{i=1}^K  J_{a_i} \sigma^z_{a_i}   } -   \sigma^x_{e_1} \right)
\nonumber \\
&& + G^{ini} \left[ \sum_{i=1}^K  J_{a_i}' \sigma^z_{a_i}  \right]
 \left( e^{ - \beta  \sigma^z_{e_2} \sum_{i=1}^K  J_{a_i}' \sigma^z_{a_i}   } -   \sigma^x_{e_2} \right)
\nonumber \\
&& +  
 \sum_{i=1}^K G_{a_i} \left[ J_{a_i} \sigma^z_{e_1}+ J_{a_i}' \sigma^z_{e_2}  \right]
 \left( e^{ - \beta \sigma^z_{a_i}  ( J_{a_i} \sigma^z_{e_1}+ J_{a_i}' \sigma^z_{e_2} )  } -   \sigma^x_{a_i} \right)
\label{HKdesbord}
\end{eqnarray}

Let us first consider the dynamical path where the spins are flipped in the order $S_{e_1},S_{a_1},S_{a_2},..,S_{a_K},S_{e_2}$.
It is convenient to introduce the following notations for the corresponding $(K+3)$ configurations
\begin{eqnarray}
C_0 &&  =\{S_{e_1}=+1,S_{a_1}=+1,S_{a_2}=+1,..,S_{a_K}=+1,S_{e_2}=+1 \}
\nonumber \\
C_1 &&  =\{S_{e_1}=-1,S_{a_1}=+1,S_{a_2}=+1,..,S_{a_K}=+1,S_{e_2}=+1 \}
\nonumber \\
C_2 &&  =\{S_{e_1}=-1,S_{a_1}=-1,S_{a_2}=+1,..,S_{a_K}=+1,S_{e_2}=+1 \}
\nonumber \\
C_p &&  =\{S_{e_1}=-1,S_{a_1}=-1,...,S_{a_{p-1}}=-1,S_{a_{p}}=+1..,S_{a_K}=+1,S_{e_2}=+1 \}
\nonumber \\
C_K &&  =\{S_{e_1}=-1,S_{a_1}=-1,...,S_{a_{K-1}}=-1,,S_{a_K}=+1,S_{e_2}=+1 \}
\nonumber \\
C_{K+1} &&  =\{S_{e_1}=-1,S_{a_1}=-1,...,S_{a_K}=-1,S_{e_2}=+1 \}
\nonumber \\
C_{K+2} &&  =\{S_{e_1}=-1,S_{a_1}=-1,...,S_{a_K}=-1,S_{e_2}=-1 \}
\label{configbord}
\end{eqnarray}
The physical meaning is that the transition from $C_0$ to $C_1$ corresponds to the entrance of a domain-wall at the boundary $e_1$, the transitions between the configurations $(C_1,C_2,..,C_{K+1})$ correspond to the displacement of this domain-wall, and finally the transition from $C_{K+1}$ to $C_{K+2}$ corresponds to the exit of the domain-wall at the boundary $e_2$.  

The two boundaries configurations are the two ferromagnetic groundstates of classical energy (Eq \ref{Uspinbord})
\begin{eqnarray}
U(C_0)=U(C_{K+2}) =  -\sum_{i=1}^K  ( J_{a_i} + J_{a_i}' )
\label{Uferrobord}
\end{eqnarray}
The intermediate configurations $C_p$ for $1 \leq p \leq K+1$ have a higher classical energy (Eq \ref{Uspinbord})
\begin{eqnarray}
U({ C_p}) = \sum_{i=1}^{p-1}  ( - J_{a_i} + J_{a_i}' ) + \sum_{i=p}^K   (  J_{a_i} - J_{a_i}' )
\label{Ucpbord}
\end{eqnarray}
So we may used the formula of Eq. \ref{Grfinal} derived in the Appendix 
for the renormalized amplitude along this dynamical path
\begin{eqnarray}
\frac{1}{G_R^{(e_1,a_1,a_2,..,a_K,e_2)} ({ C_0} , { C_{K+2}})} 
&& = e^{ - \frac{\beta}{2} \left[ U(C_0)+U(C_{K+2}) \right]}
\sum_{m=0}^{K+1} \frac{    e^{ \frac{\beta}{2} \left[  U({ C_{m}})+U({ C_{m+1}})  \right] } }
{ G\left(  C_m ,  C_{m+1} \right)  }
\label{Grbord}
\end{eqnarray}
Taking into account the local field on the spin that is flipped between two consecutive configurations,
one obtains for the two boundary flips 
\begin{eqnarray}
G\left(  C_0 ,  C_{1} \right) &&  = G^{ini} \left[\sum_{i=1}^K   J_{a_i} \right]
\nonumber \\
G\left(  C_{K+1} ,  C_{K+2} \right) &&  = G^{ini} \left[\sum_{i=1}^K   J_{a_i}' \right]
\label{Glocbord}
\end{eqnarray}
and for the intermediate flips $1 \leq m \leq K $
\begin{eqnarray}
G\left(  C_m ,  C_{m+1} \right) &&  = G_{a_m} \left[  J_{a_m}- J_{a_m}' \right]
\label{Glocbordm}
\end{eqnarray}

The renormalization formula of Eq. \ref{Grbord} for the last RG step finally reads
\begin{eqnarray}
\frac{1}{G_{Rlast}^{(e_1,a_1,a_2,..,a_K,e_2)}} 
&& = \frac{e^{ \beta \sum_{i=1}^K   J_{a_i}} }{G^{ini} \left[\sum_{i=1}^K   J_{a_i} \right]}
+ \sum_{m=1}^K  \frac{e^{ \beta \left[ 2 \sum_{i=1}^{m-1}   J_{a_i}' + J_{a_m}+  J_{a_m}' +  2 \sum_{i=m+1}^{K}   J_{a_i} \right] } }{G_{a_m} \left[  J_{a_m}- J_{a_m}' \right]}
+  \frac{e^{\beta \sum_{i=1}^K   J_{a_i}' } }{G^{ini} \left[\sum_{i=1}^K   J_{a_i}' \right]}
\label{Grbordfin}
\end{eqnarray}

\subsection{ RG rule for the bulk RG steps  $1 \leq p \leq n-1$ }

\begin{figure}[htbp]
\includegraphics[height=8cm]{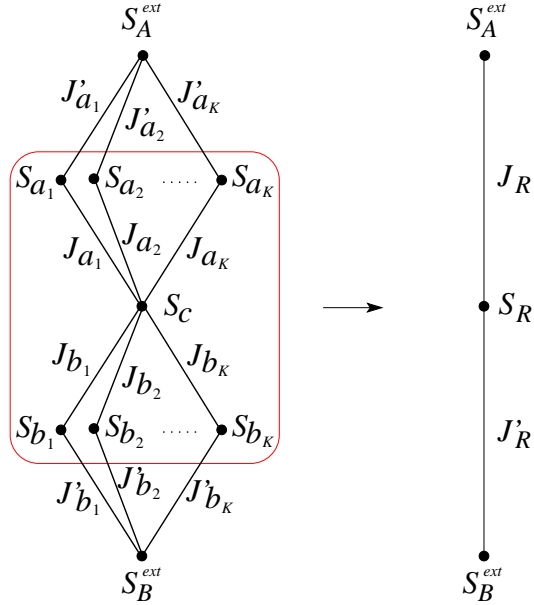}
\hspace{1cm}
\caption{ Bulk RG step for the dynamics of the disordered Ising model : notations for the quantum Hamiltonian of Eq. \ref{HKdesbulk}   }
\label{figdesbulk}
\end{figure}

The bulk RG step $p$ shown on Fig. \ref{figdesbulk}
involves two external spins $S_A^{ext} $ and $S_B^{ext} $,
and $(2K+1)$ internal spins $(S_{a_1},S_{a_2},.,S_{a_K},S_c,S_{b_1},S_{b_2},.,S_{b_K})$
with the following classical energy (Eq. \ref{Uspin})
\begin{eqnarray}
U({ C}) = -\sum_{i=1}^K  S_{a_i} ( J_{a_i} S_c+ J_{a_i}' S_A^{ext}) -\sum_{i=1}^K  S_{b_i} ( J_{b_i} S_c+ J_{b_i}' S_B^{ext})
\label{Uspinbulk}
\end{eqnarray}

The quantum Hamiltonian of Eq. \ref{Hgene} associated to the single-spin-flip dynamics reads
\begin{eqnarray}
 H_{2K+1} &&  = G^{ini} \left[ \sum_{i=1}^K ( J_{a_i} \sigma^z_{a_i}+J_{b_i} \sigma^z_{b_i} ) \right]
 \left( e^{ - \beta  \sigma^z_c \sum_{i=1}^K ( J_{a_i} \sigma^z_{a_i}+J_{b_i} \sigma^z_{b_i} )  } -   \sigma^x_c \right)
\nonumber \\
&& +  
 \sum_{i=1}^K G_{a_i} \left[ J_{a_i} \sigma^z_c+J_{a_i}'S_A^{ext} ) \right]
 \left( e^{ - \beta \sigma^z_{a_i}  ( J_{a_i} \sigma^z_c+J_{a_i}'S_A^{ext} )  } -   \sigma^x_{a_i} \right)
\nonumber \\
&& +  
 \sum_{i=1}^K  G_{b_i}  \left[  J_{b_i} \sigma^z_c+J_{b_i}' S_B^{ext} \right]
\left( e^{ - \beta \sigma^z_{b_i}  ( J_{b_i} \sigma^z_c+J_{b_i}' S_B^{ext} )  } -   \sigma^x_{b_i} \right)
\label{HKdesbulk}
\end{eqnarray}

Let us first consider the dynamical path where the spins are flipped in the order $S_{a_1},S_{a_2},..,S_{a_K},S_{c},S_{b_1},S_{b_2},..,S_{b_K}$.
More precisely, for the external Domain-Wall conditions 
\begin{eqnarray}
S_A^{ext} && =-1
\nonumber \\
S_B^{ext} && =+1
\label{dwext}
\end{eqnarray}
it is convenient to introduce the following notations for the corresponding $(2K+2)$ configurations
that describe the motion of the domain-wall 
\begin{eqnarray}
C_0 &&  =\{ S_{a_1}=+1,S_{a_2}=+1,..,S_{a_K}=+1,S_{c}=+1,S_{b_1}=+1,S_{b_2}=+1,..,S_{b_K}=+1 \}
\nonumber \\
C_1 &&  =\{ S_{a_1}=-1,S_{a_2}=+1,..,S_{a_K}=+1,S_{c}=+1,S_{b_1}=+1,S_{b_2}=+1,..,S_{b_K}=+1 \}
\nonumber \\
C_2 &&  =\{ S_{a_1}=-1,S_{a_2}=-1,S_{a_3}=+1..,S_{a_K}=+1,S_{c}=+1,S_{b_1}=+1,S_{b_2}=+1,..,S_{b_K}=+1 \}
\nonumber \\
C_K &&  =\{ S_{a_1}=-1,S_{a_2}=-1,..,S_{a_K}=-1,S_{c}=+1,S_{b_1}=+1,S_{b_2}=+1,..,S_{b_K}=+1 \}
\nonumber \\
C_{K+1} &&  =\{ S_{a_1}=-1,S_{a_2}=-1,..,S_{a_K}=-1,S_{c}=-1,S_{b_1}=+1,S_{b_2}=+1,..,S_{b_K}=+1 \}
\nonumber \\
C_{K+2} &&  =\{ S_{a_1}=-1,S_{a_2}=-1,..,S_{a_K}=-1,S_{c}=-1,S_{b_1}=-1,S_{b_2}=+1,..,S_{b_K}=+1 \}
\nonumber \\
C_{K+3} &&  =\{ S_{a_1}=-1,S_{a_2}=-1,..,S_{a_K}=-1,S_{c}=-1,S_{b_1}=-1,S_{b_2}=-1,S_{b_3}=+1,..,S_{b_K}=+1 \}
\nonumber \\
C_{2K} &&  =\{ S_{a_1}=-1,S_{a_2}=-1,..,S_{a_K}=-1,S_{c}=-1,S_{b_1}=-1,S_{b_2}=-1,..S_{b_{K-1}}=-1,S_{b_K}=+1 \}
\nonumber \\
C_{2K+1} &&  =\{ S_{a_1}=-1,S_{a_2}=-1,..,S_{a_K}=-1,S_{c}=-1,S_{b_1}=-1,S_{b_2}=-1,..S_{b_{K-1}}=-1,S_{b_K}=-1 \}
\label{configbulk}
\end{eqnarray}

The classical energy (Eq \ref{Uspinbulk}) of these configurations read for $0 \leq m \leq K$
\begin{eqnarray}
U({ C_m}) && = \sum_{i=1}^m  ( J_{a_i}  - J_{a_i}' ) -\sum_{i=m+1}^K   ( J_{a_i}  - J_{a_i}' ) -\sum_{i=1}^K   ( J_{b_i} + J_{b_i}' )
\nonumber \\
U({ C_{K+1+m}}) && = \sum_{i=1}^K   ( - J_{a_i}  - J_{a_i}' ) + \sum_{i=1}^m   ( - J_{b_i} + J_{b_i}' )
-\sum_{i=m+1}^K   ( - J_{b_i} + J_{b_i}' )
\label{Ucbulk}
\end{eqnarray}

So we may used the formula of Eq. \ref{Grfinal} derived in the Appendix 
for the renormalized amplitude along this dynamical path
\begin{eqnarray}
\frac{1}{G_R^{(a_1,a_2,..,a_K,c,b_1,b_2..,b_K)} ({ C_0} , { C_{2K+1}})} 
&& = e^{ - \frac{\beta}{2} \left[ U(C_0)+U(C_{2K+1}) \right]}
\sum_{m=0}^{2K} \frac{    e^{ \frac{\beta}{2} \left[  U({ C_{m}})+U({ C_{m+1}})  \right] } }
{ G\left(  C_m ,  C_{m+1} \right)  }
\label{Grbulk}
\end{eqnarray}
Taking into account the local field on the spin that is flipped between two consecutive configurations,
one obtains for the two boundary flips for $1 \leq m \leq K-1 $
\begin{eqnarray}
G\left(  C_m ,  C_{m+1} \right) &&  = G_{a_m} \left[  J_{a_m}- J_{a_m}' \right]
\nonumber \\
G\left(  C_{K} ,  C_{K+1} \right) &&  = G^{ini} \left[\sum_{i=1}^K   (J_{a_i}- J_{b_i}) \right]
\nonumber \\
G\left(  C_{K+m} ,  C_{K+m+1} \right) &&  = G_{b_m} \left[ J_{b_m}- J_{b_m}'  \right]
\label{Gcbulk}
\end{eqnarray}

For the renormalized spin $S_R$, the two renormalized ferromagnetic couplings read
\begin{eqnarray}
J_R && \equiv \sum_{i=1}^K   J_{a_i}'
\nonumber \\
J_R'  && \equiv \sum_{i=1}^K   J_{b_i}'
\label{JRSR}
\end{eqnarray}
and the absolute value of the renormalized local field $h_R$ in the domain-wall configurations
takes the single value
\begin{eqnarray}
\vert h_R \vert =\vert J_R - J_R' \vert = \vert \sum_{i=1}^K  ( J_{a_i}'- J_{b_i}') \vert
\label{hRSR}
\end{eqnarray}
So one does not have to renormalize a function $G[h]$ of an arbitrary local field $h$,
but only a set of three correlated variables $(J_R,J_R',G_R^{(a_1,a_2,..,a_K,c,b_1,b_2..,b_K)}[h_R= J_R - J_R'])$
representing the two ferromagnetic couplings $(J_R,J_R')$ of Eq. \ref{JRSR}
and the corresponding numerical amplitude $G_R^{(a_1,a_2,..,a_K,c,b_1,b_2..,b_K)}[h_R= J_R - J_R'] $
given by the final formula ( Eq. \ref{Grbulk} )
\begin{eqnarray}
\frac{1}{G_R^{(a_1,a_2,..,a_K,c,b_1,b_2..,b_K)} [h_R= J_R - J_R'] } 
&& = \sum_{m=1}^{K} 
\frac{e^{   \beta   \left[ \sum_{i=1}^{m-1}  ( 2J_{a_i}- J_{a_i}'- J_{b_i}') + J_{a_{m}}-J_{b_{m}}'
+  \sum_{i=m+1}^{K}  (J_{a_{i}}'- J_{b_{i}}' ) \right] }  }
{ G_{a_m}  \left[ J_{a_m}'- J_{a_m} \right]}  
  \nonumber \\
&&+ \frac{e^{   \beta \sum_{i=1}^K   (J_{a_i}+ J_{b_i} -J_{a_i}' - J_{b_i}' )  }  }
{ G^{ini}  \left[\sum_{i=1}^K   (J_{a_i}- J_{b_i})  \right]}
  \nonumber \\
&&+
 \sum_{m=1}^{K}  \frac{e^{   \beta   \left[ \sum_{i=1}^{m-1} ( J_{b_i}'  -J_{a_i}') 
+J_{b_{m}}- J_{a_{m}}' + \sum_{i=m+1}^{K}  (2J_{b_{i}}- J_{a_{i}}'- J_{b_{i}}') \right]  }  }
{ G_{b_{m}}  \left[ J_{b_{m}} -J_{b_{m}}' \right] }  
\label{Grbulkfin}
\end{eqnarray}

\subsection{ Analysis of the random ferromagnetic chain $(K=1)$  }

For the random ferromagnetic chain corresponding to the branching ratio $K=1$,
the renormalization formula for the last step (Eq. \ref{Grbordfin}) becomes 
\begin{eqnarray}
\frac{1}{G_{last}^{(e_1,a_1,e_2)} }
&& = \frac{e^{ \beta    J_{a_1}} }{G^{ini} \left[  J_{a_1} \right]}
+   \frac{e^{ \beta \left[  J_{a_1}+  J_{a_1}'  \right] } }{G_{a_1} \left[  J_{a_1}- J_{a_1}' \right]}
+  \frac{e^{\beta    J_{a_1}' } }{G^{ini} \left[  J_{a_1}' \right]}
\label{GrbordfinK1}
\end{eqnarray}
For the bulk, we may rewrite Eq. \ref{Grbulkfin} as
\begin{eqnarray}
\frac{e^{ \beta \left[  J_{a_1}'+  J_{b_1}'  \right] } }{G_R [ J_{a_1}' - J_{b_1}'] }
&& = 
\frac{e^{   \beta   \left[  J_{a_{1 }}+J_{a_{1 }}' \right] }  }
{ G_{a_1 }  \left[ J_{a_1 }'- J_{a_1 } \right]}  
+ \frac{e^{   \beta    (J_{a_1}+ J_{b_1}  )  }  }
{ G^{ini}  \left[ J_{a_1}- J_{b_1}  \right]}
+
 \frac{e^{   \beta   \left[ 
J_{b_{1 }}+ J_{b_{1 }}' \right]  }  }
{ G_{b_{1 }}  \left[ J_{b_{1 }} -J_{b_{1 }}' \right] }  
\label{GrbulkfinK1}
\end{eqnarray}
to make clearer that the combination
 $\frac{e^{ \beta \left[  J_{a_1}'+  J_{b_1}'  \right] } }{G_R [ J_{a_1}' - J_{b_1}'] } $  has a simple renormalization rule.
By iteration, we finally obtain that the final amplitude $G_{last}^{(e_1,e_2)}(L_n=2^n) $
for a system of size $L_n=2^n$ of $n$ generations 
reads in terms of the initial function $G^{ini}[h]$  (with the notations $J_{-1,0}=0=J_{L_n,L_n+1} $)
\begin{eqnarray}
\frac{1}{G_{last}^{(e_1,e_2)} (L_n=2^n) }
&& = \sum_{i=1}^{L_n=2^n}  \frac{e^{ \beta  (J_{i-1,i}+J_{i,i+1})  } }{G^{ini} \left[ J_{i-1,i}-J_{i,i+1}  \right]}
\label{GlastK1}
\end{eqnarray}
when the domain-wall enters by the boundary $e_1$ and exits by the boundary $e_2$

We should now take into account the other case where the domain-wall enters by the boundary $e_2$ and exits by the boundary $e_1$, which actually gives the same contribution
\begin{eqnarray}
G_{last}^{(e_2,e_1)} (L_n=2^n) = G_{last}^{(e_1,e_2)} (L_n=2^n)
\label{GlastK1bis}
\end{eqnarray}
The total amplitude $G_{last}(L_n=2^n)$ is the sum of these two contributions
\begin{eqnarray}
G_{last} (L_n=2^n)=G_{last}^{(e_1,e_2)} (L_n=2^n)+G_{last}^{(e_2,e_1)} (L_n=2^n) = 2 G_{last}^{(e_1,e_2)} (L_n=2^n)
\label{GlastK1sum}
\end{eqnarray}
so that the final result reads
\begin{eqnarray}
\frac{1}{G_{last} (L_n=2^n) }
 = \frac{1}{2} 
\sum_{i=1}^{L_n=2^n}  \frac{e^{ \beta  (J_{i-1,i}+J_{i,i+1})  } }{G^{ini} \left[ J_{i-1,i}-J_{i,i+1}  \right]}
\label{GlastK1final}
\end{eqnarray}
in agreement with the results of Eq. (104) obtained in our previous work \cite{us_rgdyn}
via the Boundary Renormalization procedure.
This agreement shows the validity of the Bulk Renormalization procedure within the domain-wall approximation.

\subsection{ Dynamical barriers for branching ratio $K>1$  }

For $K>1$, we have to compare the various dynamical paths that display different dynamical barriers
as a consequence of the disorder.

As explained before Eq. \ref{Grbulkfin}, the important quantity is the numerical amplitude $G_R[h_R=J_R-J_R']$
which is correlated with the two renormalized ferromagnetic couplings $(J_R,J_R')$.
It is thus convenient to introduce the corresponding dynamical barrier ${\cal B}_R^{(J_R,J_R')}$
defined by the low temperature behavior 
\begin{eqnarray}
G_{R} \left[  J_{R}- J_{R}' \right] \opsimeq_{\beta \to \infty} e^{- \beta {\cal B}_R^{(J_R,J_R')} }
\label{gbdyn}
\end{eqnarray}
where the notation ${\cal B}_R^{(J_R,J_R')}$ has been chosen to remind that this barrier is
{\it correlated } with the two couplings $(J_R,J_R') $.

Let us now focus on the Glauber dynamics of Eq. \ref{glauber} with the following low-temperature behavior
\begin{eqnarray}
G^{ini}_{Glauber} [h] = \frac{1}
{ 2 \cosh \left( \beta  h \right) } \opsimeq_{\beta \to \infty} e^{ - \beta \vert h \vert }
\label{glaubersmallT}
\end{eqnarray}

\subsubsection{ Optimization of the dynamical path  for the last RG step $p=n$  }

In terms of dynamical barriers, 
 Eq. \ref{Grbordfin} with Eq. \ref{glaubersmallT}
yields that the final dynamical barrier $ {\cal B}_{last}^{(a_1,a_2,..,a_K)}$
associated to the given dynamical path $(a_1,a_2,..,a_K)$ reads
\begin{eqnarray}
&&  {\cal B}_{last}^{(a_1,a_2,..,a_K)} 
= \max \left[ 2 \sum_{i=1}^K   J_{a_i} ; 2 \sum_{i=1}^K   J_{a_i}' ;
\max_{1 \leq m \leq K} \left( {\cal B}_{a_m}^{(J_{a_m},J_{a_m}')} 
+ 2 \sum_{i=1}^{m-1}   J_{a_i}' + J_{a_m}+  J_{a_m}' +  2 \sum_{i=m+1}^{K}   J_{a_i} \right)
 \right]
\label{Bpathbord}
\end{eqnarray}

We now have to consider the $K!$ possible dynamical paths :
for a given permutation $\pi$ of the $K$ renormalized spins, 
the dynamical barrier ${\cal B}_{last}^{(a_{\pi(1)},a_{\pi(2)},..,a_{\pi(K)})}$
 associated to the path $(a_{\pi(1)},a_{\pi(2)},..,a_{\pi(K)})$
reads by adapting Eq \ref{Bpathbord}
 \begin{eqnarray}
&& {\cal B}_{last}^{(a_{\pi(1)},..,a_{\pi(K)}) }
 \nonumber \\ &&  = \max \left[ 2 \sum_{i=1}^K   J_{a_i} ; 2 \sum_{i=1}^K   J_{a_i}' ;
\max_{1 \leq m \leq K} \left( {\cal B}_{a_{\pi(m)}}^{(J_{a_{\pi(m)}},J_{a_{\pi(m)}}')} 
+ 2 \sum_{i=1}^{m-1}   J_{a_{\pi(i)}}' + J_{a_{\pi(m)}}+  J_{a_{\pi(m)}}' +  2 \sum_{i=m+1}^{K}   J_{a_{\pi(i)}} \right)
 \right]
\label{brpathpi}
\end{eqnarray}

We now have to choose the dynamical path, i.e. the
permutation $\pi$ leading to the smallest barrier. So
the final renormalized barrier ${\cal B}_{last} $
is given by the minimum of Eq. \ref{brpathpi}
over the $K!$ possible permutations
 \begin{eqnarray}
&& {\cal B}_{last}  \equiv \min_{\pi} \left(  {\cal B}_{last}^{(a_{\pi(1)},..,a_{\pi(K)}) }  \right)
 \\ &&  = 
\min_{\pi} \left(  \max \left[ 2 \sum_{i=1}^K   J_{a_i} ; 2 \sum_{i=1}^K   J_{a_i}' ;
\max_{1 \leq m \leq K} \left( {\cal B}_{a_{\pi(m)}}^{(J_{a_{\pi(m)}},J_{a_{\pi(m)}}')} 
+ 2 \sum_{i=1}^{m-1}   J_{a_{\pi(i)}}' + J_{a_{\pi(m)}}+  J_{a_{\pi(m)}}' +  2 \sum_{i=m+1}^{K}   J_{a_{\pi(i)}} \right)
 \right]
 \right)
\nonumber
\label{btotbord}
\end{eqnarray}

\subsubsection{ Optimization of the dynamical path  for the bulk RG steps $1 \leq p \leq n-1$  }

Similarly for the bulk RG step, Eq. \ref{Grbulkfin} 
yields that the renormalized dynamical barrier $ {\cal B}_{R}^{(a_1,a_2,..,a_K,c,b_1,b_2..,b_K) }$
correlated with the renormalized couplings (Eq. \ref{JRSR})
\begin{eqnarray}
J_R && \equiv \sum_{i=1}^K   J_{a_i}'
\nonumber \\
J_R'  && \equiv \sum_{i=1}^K   J_{b_i}'
\label{JRSRbulk}
\end{eqnarray}
and associated to the dynamical path $(a_1,a_2,..,a_K,c,b_1,b_2..,b_K) $
 reads
\begin{eqnarray}
{\cal B}_{R}^{(a_1,a_2,..,a_K,c,b_1,b_2..,b_K) }
= \max ( && \max_{1 \leq m \leq K} 
 \left[ {\cal B}_{a_m}^{(J_{a_m},J_{a_m}')} + \sum_{i=1}^{m-1}  ( 2J_{a_i}- J_{a_i}'- J_{b_i}') + J_{a_{m}}-J_{b_{m}}'
+  \sum_{i=m+1}^{K}  (J_{a_{i}}'- J_{b_{i}}' ) \right]
;\nonumber
 \\ && \big\vert \sum_{i=1}^K   (J_{a_i}- J_{b_i}) \big\vert + \sum_{i=1}^K   (J_{a_i}+ J_{b_i} -J_{a_i}' - J_{b_i}' )
;\label{Brpathbulk}
 \\ && \max_{1 \leq m \leq K}   \left[ {\cal B}_{b_m}^{(J_{b_m},J_{b_m}')} + \sum_{i=1}^{m-1} ( J_{b_i}'  -J_{a_i}') 
+J_{b_{m}}- J_{a_{m}}' + \sum_{i=m+1}^{K}  (2J_{b_{i}}- J_{a_{i}}'- J_{b_{i}}') \right]
)
\nonumber
\end{eqnarray}

We now have to consider the $(K!)^2$ possible dynamical paths :
for a given permutations $(\pi_a)$ of the $a_m$ renormalized spins, 
and for a given permutations $(\pi_b)$ of the $b_m$ renormalized spins,
the dynamical barrier ${\cal B}_{R}^{(a_{\pi_a(1)},a_{\pi_a(2)},..,a_{\pi_a(K)},c,b_{\pi_b(1)},b_{\pi_b(2)},..,b_{\pi_b(K)})}$
 associated to the path $(a_{\pi_a(1)},a_{\pi_a(2)},..,a_{\pi_a(K)},c,b_{\pi_b(1)},b_{\pi_b(2)},..,b_{\pi_b(K)}) $
reads by adapting Eq \ref{Brpathbulk}
\begin{eqnarray}
&& {\cal B}_{R}^{ (a_{\pi_a(1)},a_{\pi_a(2)},..,a_{\pi_a(K)},c,b_{\pi_b(1)},b_{\pi_b(2)},..,b_{\pi_b(K)}) }= \max (
\nonumber \\
 && \max_{1 \leq m \leq K} 
 \left[ {\cal B}_{a_{\pi_a(m)}}^{(J_{a_{\pi_a(m)}},J_{a_{\pi_a(m)}}')} + \sum_{i=1}^{m-1}  ( 2J_{a_{\pi_a(i)}}- J_{a_{\pi_a(i)}}'- J_{b_{\pi_b(i)}}') + J_{a_{\pi_a(m)}}-J_{b_{\pi_b(m)}}'
+  \sum_{i=m+1}^{K}  (J_{a_{\pi_a(i)}}'- J_{b_{\pi_b(i)}}' ) \right]
;\nonumber
 \\ && \big\vert \sum_{i=1}^K   (J_{a_i}- J_{b_i}) \big\vert + \sum_{i=1}^K   (J_{a_i}+ J_{b_i} -J_{a_i}' - J_{b_i}' )
;
 \\ && \max_{1 \leq m \leq K}   \left[ {\cal B}_{b_{\pi_b(m)}}^{(J_{b_{\pi_b(m)}},J_{b_{\pi_b(m)}}')} + \sum_{i=1}^{m-1} ( J_{b_{\pi_b(i)}}'  -J_{a_{\pi_a(i)}}') 
+J_{b_{\pi_b(m)}}- J_{a_{\pi_a(m)}}' + \sum_{i=m+1}^{K}  (2J_{b_{\pi_b(i)}}- J_{a_{\pi_a(i)}}'- J_{b_{\pi_b(i)}}') \right]
)
\nonumber
\label{Brpathbulkpi}
\end{eqnarray}

We now have to choose the dynamical path, i.e. the
permutations $(\pi_a,\pi_b)$ leading to the smallest barrier. So
the final renormalized barrier ${\cal B}_{R}^{(J_R,J_R')} $
is given by the minimum of Eq. \ref{Brpathbulkpi}
over the $K!$ permutations $(\pi_a)$ and over the $K!$ permutations $(\pi_b)$
 \begin{eqnarray}
 {\cal B}_{R}^{(J_R,J_R')}  \equiv \min_{\pi_a,\pi_b} \left( {\cal B}_{R}^{ (a_{\pi_a(1)},a_{\pi_a(2)},..,a_{\pi_a(K)},c,b_{\pi_b(1)},b_{\pi_b(2)},..,b_{\pi_b(K)}) } \right)
\label{btotbulk}
\end{eqnarray}
To see more clearly the structure, let us now focus on the case $K=2$.

\subsection{ Case $K=2$ corresponding to the fractal dimension $d_f=2$ }

\subsubsection{ Last step RG  rule for dynamical barriers when $K=2$  }

For $K=2$, Eq. \ref{Bpathbord} involves a maximum over four terms
\begin{eqnarray}
 {\cal B}_{last}^{(a_1,a_2)} && = \max \left[ 2 (J_{a_1}+J_{a_2})  ; 2 (J_{a_1}'+J_{a_2}');
  {\cal B}_{a_1}^{(J_{a_1},J_{a_1}')} 
+ J_{a_1}+  J_{a_1}' +  2    J_{a_2} ;
  {\cal B}_{a_2}^{(J_{a_2},J_{a_2}')} 
+ 2    J_{a_1}' + J_{a_2}+  J_{a_2}' 
 \right]
\label{BpathbordK2}
\end{eqnarray}
and  Eq. \ref{btotbord} involves the minimum over two permutations
 \begin{eqnarray}
 {\cal B}_{last} = 
\min(&& \! \! \! \! \! \max \left[ 2 (J_{a_1}+J_{a_2})  ; 2 (J_{a_1}'+J_{a_2}');
  {\cal B}_{a_1}^{(J_{a_1},J_{a_1}')} 
+ J_{a_1}+  J_{a_1}' +  2    J_{a_2} ;
  {\cal B}_{a_2}^{(J_{a_2},J_{a_2}')} 
+ 2    J_{a_1}' + J_{a_2}+  J_{a_2}' 
 \right] ;
\nonumber
 \\ && \! \! \! \! \!
\max \left[ 2 (J_{a_1}+J_{a_2})  ; 2 (J_{a_1}'+J_{a_2}');
  {\cal B}_{a_2}^{(J_{a_2},J_{a_2}')} 
+ J_{a_2}+  J_{a_2}' +  2    J_{a_1} ;
  {\cal B}_{a_1}^{(J_{a_1},J_{a_1}')} 
+ 2    J_{a_2}' + J_{a_1}+  J_{a_1}' 
 \right]
 )
\label{btotbordK2}
\end{eqnarray}

\subsubsection{ Bulk RG  rules for dynamical barriers when $K=2$  }

Eq \ref{Brpathbulk} involves a maximum over five terms
\begin{eqnarray}
{\cal B}_{R}^{(a_1,a_2,c,b_1,b_2) }
= \max [ && {\cal B}_{a_1}^{(J_{a_1},J_{a_1}')} +  J_{a_{1}}-J_{b_{1}}'
+    (J_{a_{2}}'- J_{b_{2}}' ) 
;\nonumber
 \\ &&
{\cal B}_{a_2}^{(J_{a_2},J_{a_2}')} + ( 2J_{a_1}- J_{a_1}'- J_{b_1}')
+ J_{a_{2}}-J_{b_{2}}' ;
\nonumber
 \\ &&
 \big\vert \sum_{i=1}^2   (J_{a_i}- J_{b_i}) \big\vert + \sum_{i=1}^2   (J_{a_i}+ J_{b_i} -J_{a_i}' - J_{b_i}' )
;
\label{BrpathbulkK2} \\ &&  {\cal B}_{b_1}^{(J_{b_1},J_{b_1}')} 
+J_{b_{1}}- J_{a_{1}}' + (2J_{b_{2}}- J_{a_{2}}'- J_{b_{2}}') 
) ;
\nonumber
\\ &&  {\cal B}_{b_2}^{(J_{b_2},J_{b_2}')} +  ( J_{b_1}'  -J_{a_1}') 
+J_{b_{2}}- J_{a_{2}}' 
]
\nonumber
\end{eqnarray}

and Eq. \ref{btotbulk} involves the minimum over four terms
 \begin{eqnarray}
 {\cal B}_{R}^{(J_R,J_R')} = \min \left[ {\cal B}_{R}^{(a_1,a_2,c,b_1,b_2) }; {\cal B}_{R}^{(a_2,a_1,c,b_1,b_2) } ;
{\cal B}_{R}^{(a_1,a_2,c,b_2,b_1) } ;  {\cal B}_{R}^{(a_2,a_1,c,b_2,b_1) } \right]
\label{btotbulkK2}
\end{eqnarray}

\subsubsection{ Numerical results obtained via the pool method  }

\begin{figure}[htbp]
 \includegraphics[height=6cm]{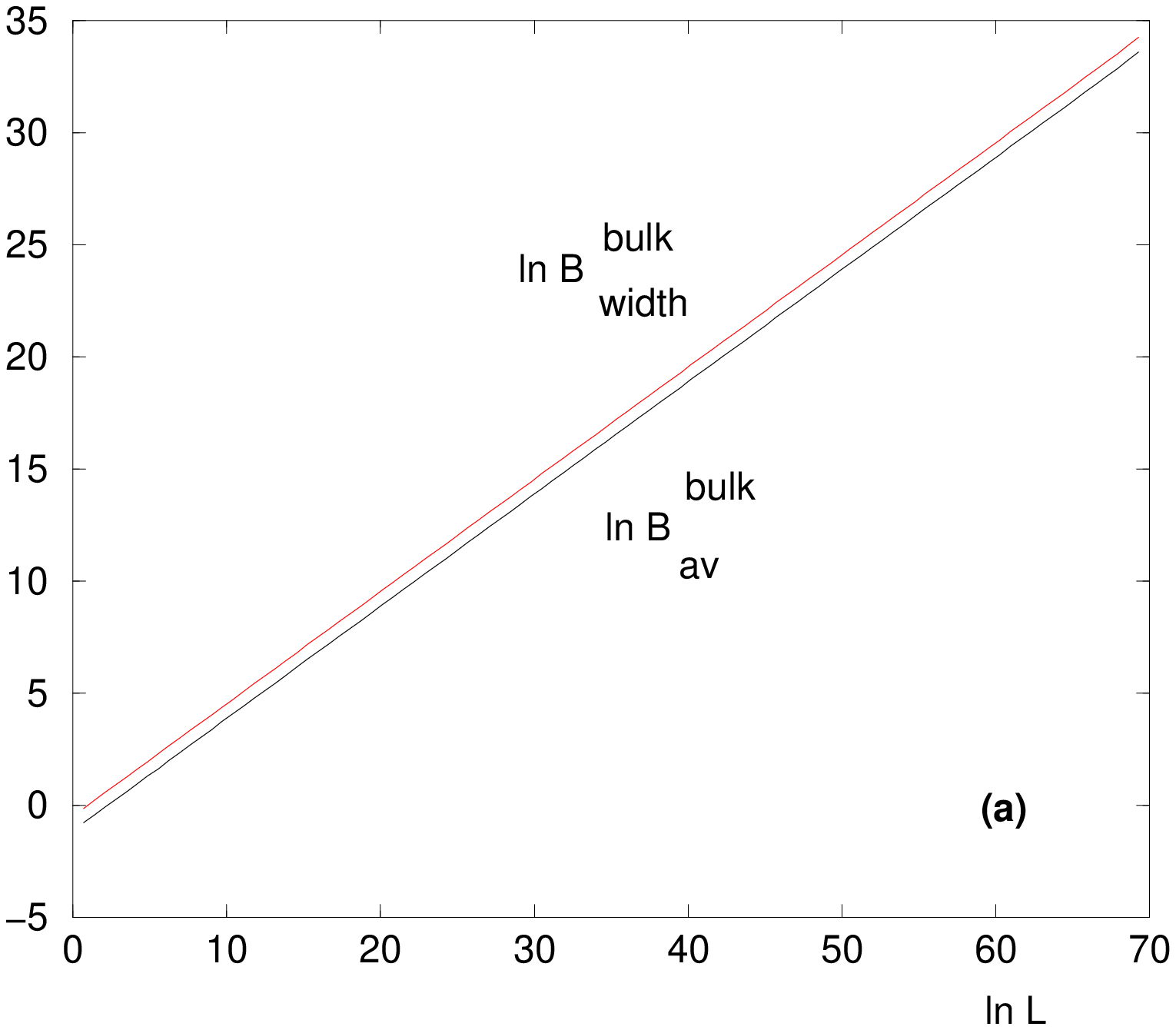}
\hspace{1cm}
 \includegraphics[height=6cm]{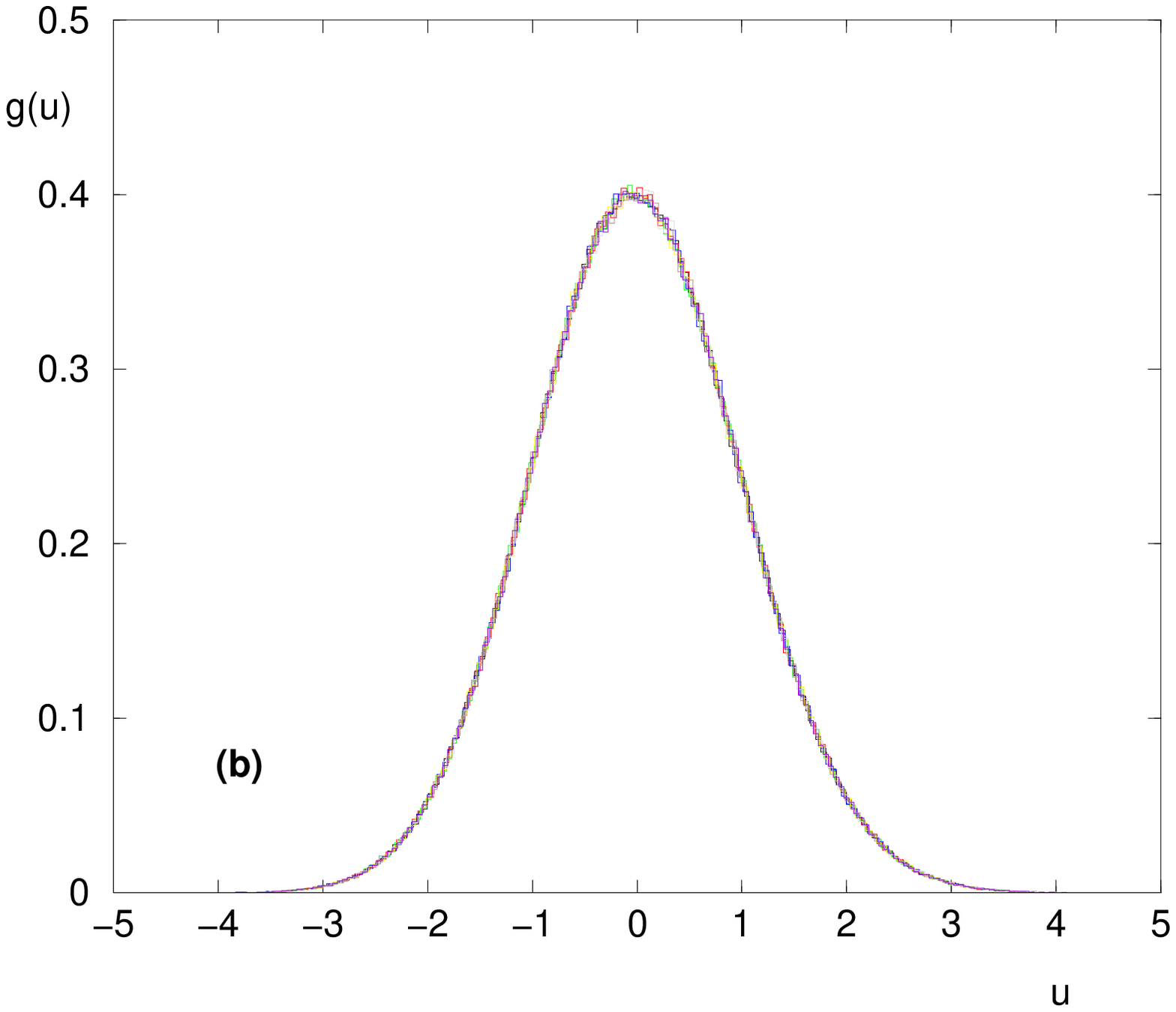}
\caption{ Glauber dynamics of the random ferromagnetic model 
defined on the diamond hierarchical lattice of branching ratio $K=2$ 
corresponding to the fractal dimension $d_f=2$ :
(a) Log-log plot of the averaged-value $B^{bulk}_{av}(n)$ and of the width $B^{bulk}_{width}(n)$ of the probability distribution $P_n(B)$
of bulk dynamical barriers at generation $n$ corresponding to the length $L_n=2^n$ :
the slope yields the dynamical exponent $\psi \simeq 0.5$
(b) The distribution of the rescaled barrier $u \equiv (\frac{B-B^{bulk}_{av}(n)}{B^{bulk}_{width}(n)})$
is the Gaussian distribution $g(u)$ of Eq. \ref{gaussb}. }
\label{figk2}
\end{figure}

The pool method is very useful to study 
renormalization rules for disordered models defined 
on trees \cite{abou,bell,us_cayley,us_rgdyntree}
and on hierarchical lattices \cite{us_potts,poolspinglass,Coo_Der,us_poly}.
The idea of the pool method is the following :  
at each generation, one keeps the same number $M_{pool}$
of random variables to represent probability distributions.
Within our present framework, the joint probability distribution $P_p({\cal B},J,J')$ 
of the dynamical barrier ${\cal B}$ and of the two renormalized couplings of a renormalized spin at
 generation $n$ will be represented by a pool of $M_{pool}=10^6$ triplets $({\cal B}_i,J_i,J_i')$.
To construct a new triplet $({\cal B}_R,J_R,J_R')$ of generation $(p+1)$, one draws $(2 K)$ triplets
$({\cal B}_i,J_i,J_i')$ within the pool of generation $p$ and apply the rule of Eqs \ref{BrpathbulkK2} 
and \ref{btotbulkK2}.
At the last RG step $p=n$, one draws instead $K$ triplets
$({\cal B}_i,J_i,J_i')$ within the pool of generation $(n-1)$ and apply the rule of Eqs \ref{BpathbordK2}
and \ref{btotbordK2}
to obtain the final barrier ${\cal B}_n^{last}$ between the two ferromagnetic groundstates of the whole sample
of length $L_n=2^n$ (eq \ref{lndiamond}).

For the Glauber dynamics satisfying Eq. \ref{glaubersmallT},
the initial condition at generation $n=0$
 reads in terms of the initial disorder distribution $\rho(J)$
of the ferromagnetic coupling reads
 \begin{eqnarray}
P^{bulk}_{n=0}({\cal B},J,J') = \rho(J) \rho(J' ) \delta ({\cal B} -\vert J-J' \vert  )
\label{inipoolgaluber}
\end{eqnarray}
We have chosen the box distribution of width $\Delta=1$
 \begin{eqnarray}
\rho(J)= \theta(1 \leq J \leq 2)
\label{rhobox}
\end{eqnarray}

On Fig. \ref{figk2}, we present our numerical results concerning the probability distribution
 of the bulk dynamical barrier ${\cal B} $ at generation $n \leq 100$
 \begin{eqnarray}
P^{bulk}_{n}({\cal B}) = \int dJ \int dJ' P^{bulk}_{n}({\cal B},J,J')
\label{pbulk}
\end{eqnarray}
As explained in previous sections, this bulk dynamical barrier ${\cal B} $ characterizes
the dynamics of a domain-wall crossing the system after its creation.
We find that both the averaged value
 \begin{eqnarray}
B^{bulk}_{av}(n) \equiv \int d{\cal B} \ {\cal B}  P^{bulk}_{n}({\cal B}) 
\label{avpbulk}
\end{eqnarray}
and the width
 \begin{eqnarray}
B^{bulk}_{width}(n) \equiv  \left( \int d{\cal B} \ {\cal B}^2  P^{bulk}_{n}({\cal B}) - (B^{bulk}_{av}(n))^2 \right)^{\frac{1}{2}}
\label{widthpbulk}
\end{eqnarray}
grow with the same power-law of the length $L_n=2^n$
 \begin{eqnarray}
B^{bulk}_{av}(n) && \propto  L_n^{\psi}
\nonumber \\
B^{bulk}_{width}(n) && \propto  L_n^{\psi}
\label{defpsi}
\end{eqnarray}
with the dynamical exponent (see Fig. \ref{figk2} (a))
 \begin{eqnarray}
\psi \simeq 0.5
\label{respsi}
\end{eqnarray}
As shown on Fig. \ref{figk2} (b), the corresponding rescaled barrier
 \begin{eqnarray}
u \equiv \frac{B-B^{bulk}_{av}(n)}{B^{bulk}_{width}(n)}
\label{rescalb}
\end{eqnarray}
follows the Gaussian distribution
 \begin{eqnarray}
g(u) = \frac{1}{\sqrt{2 \pi}} e^{- \frac{u^2}{2}}
\label{gaussb}
\end{eqnarray}

We have also computed the probability distribution of the last barrier $B^{last}$ as a function of the generation $n$.
As explained in previous sections, this last dynamical barrier ${\cal B} $ characterizes
the dynamics where a domain-wall is created near a boundary and then crosses the system.
We find that the averaged value grows linearly
 \begin{eqnarray}
B^{last}_{av}(n) && \propto  L_n
\label{blastav}
\end{eqnarray}
as expected from the energy cost $L^{d_s}$ of the creation of an interface of
dimension $d_s=d_f-1=1$ in a space of dimension $d_f=2$, in agreement with the result of
 Eq. \ref{Beqpursolu} corresponding to the pure case.
The width around this averaged value is found to scale as the width of the bulk barrier of Eq. \ref{defpsi}
 \begin{eqnarray}
B^{last}_{width}(n)  \propto  L_n^{\psi} \ \ {\rm with } \ \ \psi \simeq 0.5
\label{bulkpsi}
\end{eqnarray}

 The result $\psi=1/2$ for the dynamical exponent $\psi$ is in agreement with the conjecture $\psi=d_s/2$ proposed in
our previous work \cite{us_conjecturepsi}. In particular, $\psi=1/2$ is clearly different from the droplet exponent $\theta \simeq 0.299$ involved in the statics of the random ferromagnet on the diamond lattice \cite{us_tails}, which coincides with the Directed Polymer droplet exponent $\theta_{DP} \simeq 0.299$ \cite{Der_Gri}, since the optimization of the position of the interface in the random ferromagnet corresponds to the optimization of the position of a directed polymer in a random medium. We refer to Ref. \cite{us_conjecturepsi} for a detailed discussion of the physical meaning of the conjecture $\psi=d_s/2$ with respect to other alternative proposals.

 \section{Conclusion}

\label{sec_conclusion}

To characterize the stochastic single-spin-flip dynamics near zero-temperature of the pure and random ferromagnetic Ising model on the hierarchical diamond lattice of branching ratio $K$ with fractal dimension $d_f=(\ln (2K))/\ln 2$, we have adapted the Real Space Renormalization procedure introduced in our previous work \cite{us_rgdyn}.

 For the pure Ising model, we have obtained that the equilibrium time behaves as 
 \begin{eqnarray}
t_{eq}(L) \sim L^{\alpha} e^{\beta 2J L^{d_s}}
\label{conteqpure}
\end{eqnarray}
 where $d_s=d_f-1$ is the expected interface dimension. We have computed the prefactor exponent $\alpha$ as a function of $K$. 

For the random ferromagnetic Ising model, we have derived the renormalization rules for dynamical barriers $B_{eq}(L) \equiv (\ln t_{eq}/\beta)$ near zero temperature. For the fractal dimension $d_f=2$
(corresponding to the branching ratio $K=2$), we have studied numerically these renormalization rules via the pool method to obtain
 \begin{eqnarray}
B_{eq}(L) \sim c L+L^{1/2} u
\label{conBrandom}
\end{eqnarray}
where $u$ is a $O(1)$ Gaussian random variable of non-zero mean.
The non-random term scaling as $L$ corresponds
 to the energy-cost of the creation of an interface of dimension $d_s=d_f-1$
as in the pure case of Eq. \ref{conteqpure}. 
The dynamical exponent $\psi$ governing the fluctuation part 
characterizes the barriers for the motion of a domain-wall after its creation.
 The result $\psi=1/2$ is in agreement with the conjecture $\psi=d_s/2$ proposed in \cite{us_conjecturepsi}. 
In particular, the dynamical exponent $\psi=1/2$ is clearly different from the droplet exponent $\theta \simeq 0.299$ involved in the statics of the random ferromagnet on the same lattice \cite{us_tails}.

 \appendix

\section{  Renormalization rule for an effective one-dimensional dynamics   }

\label{sec_app}

\subsection{ First excited quantum state }

Since the exact groundstate $ \vert  \psi_0 >$ of zero energy $E_0=0$ is exactly known to be given by Eq. \ref{psi0},
it is natural to look for the first excited state through an amplitude $A({  C})$
\begin{eqnarray}
\vert  \psi_1 > = \sum_{ C} A({  C})  \frac{ e^{- \frac{\beta}{2} U({ C}) }}{\sqrt Z}
\vert {  C}  >
\label{psi1}
\end{eqnarray}
Then the eigenequation for the quantum Hamiltonian of Eq. \ref{tight}
\begin{eqnarray}
0= ({ H} -E_1 ) \vert  \psi_1 >
\label{eigenpsi1}
\end{eqnarray}
can be rewritten for the amplitude $ A({  C}) $ as 
\begin{eqnarray}
 \left[W_{out}(C)-E_1 \right] A(C) = \sum_{C'} W(C \to C') A(C')
\label{eqAgene}
\end{eqnarray}

\subsection{ Explicit first non-vanishing energy $E_1$ for  an effective one-dimensional dynamics }

Let us consider an effective one-dimensional dynamics between
 configurations $(C_0;C_1,C_2,..,C_n,C_{n+1})$
described by the system (Eq \ref{eqAgene}) for $1 \leq i \leq n$
\begin{eqnarray}
0 && = \left[W(C_i \to C_{i-1})+W(C_i \to C_{i+1} )-E_1 \right] A(C_i)
 - W(C_i \to C_{i-1}) A(C_{i-1})- W(C_i \to C_{i+1}) A(C_{i+1})
\label{eqAgenebulk}
\end{eqnarray}
and by the two boundary equations for $i=0$ and $i=n+1$
\begin{eqnarray}
0 && = \left[W(C_0 \to C_{1} )-E_1 \right] A(C_0) - W(C_0 \to C_{1}) A(C_{1})
\nonumber \\
0 && = \left[W(C_{n+1} \to C_{n})-E_1 \right] A(C_{n+1})
 - W(C_{n+1} \to C_{n}) A(C_{n})
\label{eqAgenebord}
\end{eqnarray}

Let us assume that the intermediate configurations $C_i$ for $i=1,2,..,n$
have higher classical energies $U(C_i)$ with respect to the
two boundary configurations $C_0$ and $C_{n+1}$.
Then the groundstate $ \vert  \psi_0 >$ of zero energy $E_0=0$ of Eq. \ref{psi0}
can be approximated at low temperature by its two leading components
\begin{eqnarray}
\vert \psi_0 > \opsimeq_{\beta \to +\infty} \frac{1 }{\sqrt{e^{- \beta U({ C_0 }) }+ e^{- \beta U({ C_{n+1} }) }}}
 \left(  e^{- \frac{\beta}{2} U({ C_0 }) }\vert {  C_0}  >
+ e^{- \frac{\beta}{2} U({ C_{n+1} }) }\vert {  C_{n+1} }  > \right)
\label{psizeroleading}
\end{eqnarray}
Then the small energy $E_1$ can be neglected in Eq. \ref{eqAgenebulk}
to become for $1 \leq i \leq n$
\begin{eqnarray}
 A(C_i) = p_-(C_i)  A(C_{i-1})+ p_+(C_i)  A(C_{i+1})
\label{eqA}
\end{eqnarray}
with the notations
\begin{eqnarray}
p_-(C_i) && \equiv \frac{ W(C_i \to C_{i-1}) }{W(C_i \to C_{i-1})+  W(C_i \to C_{i+1})}
\nonumber \\
p_+(C_i) && \equiv \frac{ W(C_i \to C_{i+1}) }{W(C_i \to C_{i-1})+  W(C_i \to C_{i+1})}=1-p_-(C_i)
\label{defppm}
\end{eqnarray}
The leading components of the first excited state at low temperature
\begin{eqnarray}
\vert \psi_1 > \opsimeq_{\beta \to +\infty} \psi_1(C_0) \vert {  C_0}  >
+ \psi_1(C_{n+1}) \vert {  C_{n+1} }  >
\label{psi1leading}
\end{eqnarray}
are then fixed by orthogonality with the groundstate of Eq. \ref{psizeroleading}
\begin{eqnarray}
\psi_1(C_0) &&\opsimeq_{\beta \to +\infty}  - \frac{ e^{- \frac{\beta}{2} U({ C_{n+1} }) } }
{ \sqrt{ e^{- \beta U({ C_0 }) }+ e^{- \beta U({ C_{n+1} }) } } }
\nonumber \\
\psi_1(C_{n+1}) && \opsimeq_{\beta \to +\infty}     \frac{e^{- \frac{\beta}{2} U({ C_0 }) }  }
{\sqrt{ e^{- \beta U({ C_0 }) }+ e^{- \beta U({ C_{n+1} }) }} }
\label{psi1orthog}
\end{eqnarray}
 so that the amplitude $A(C)$ satisfies the boundary conditions
\begin{eqnarray}
A(C_0) &&  \opsimeq_{\beta \to +\infty} - e^{- \frac{\beta}{2} \left[U(C_{n+1})- U(C_0)\right] }
\nonumber \\
A(C_{n+1}) && \opsimeq_{\beta \to +\infty}  e^{ \frac{\beta}{2} \left[U(C_{n+1})- U(C_0)\right] }
\label{bcampli}
\end{eqnarray}

The solution of Eq. \ref{eqA} with the boundary conditions of Eq. \ref{bcampli} 
can be obtained by recurrence \cite{rec1d} and reads
\begin{eqnarray}
 A(C_i) =  A(C_0) \frac{R_0(i,n)}{R_0(0,n)}+ A(C_{n+1}) \frac{R_{n+1}(1,i)}{R_{n+1}(1,n+1)}
\label{eqAsol}
\end{eqnarray}
in terms of the Kesten variables \cite{kestenv} 
\begin{eqnarray}
R_0(n+1,n) && =0  \\
R_0(n,n) && =1 \nonumber \\
R_0(k \leq n-1 , n) && =1+\sum_{m=k+1}^{n} \prod_{i=m}^{n} \frac{ p_+(i)}{ p_-(i)}
\nonumber \\
R_0(0,n) && = 1+\sum_{m=1}^{n} \prod_{i=m}^{n} \frac{ p_+(i)}{ p_-(i)}
 = 1+ \frac{p_+(n)}{p_-(n)} 
+ ... +
 \frac{p_+(n) p_+(n-1)...p_+(1)}{p_-(n) p_-(n-1)...p_-(1)}
\nonumber 
\label{kestenzero}
\end{eqnarray}
and
\begin{eqnarray}
R_{n+1}(1,0) && =0 \nonumber \\
R_{n+1}(1,1) && =1 \nonumber \\
R_{n+1}(1,k \geq 2) && =1+\sum_{m=1}^{k-1} \prod_{i=1}^{m} \frac{ p_-(i)}{ p_+(i)}
\nonumber \\
R_{n+1}(1,n+1) && = 1+\sum_{m=1}^{n} \prod_{i=1}^{m} \frac{ p_-(i)}{ p_+(i)}
 = 1+ \frac{p_-(1)}{p_+(1)} + \frac{p_-(1)p_-(2)}{p_+(1)p_+(2)} +
 ... + \frac{p_-(1) p_-(2)...p_-(n)}{p_+(1) p_+(2)...p_+(n)}
\label{kestennp1}
\end{eqnarray}

The energy $E_1$ can be now computed from Eq. \ref{eqAgenebord} at the boundary $C_0$
(or equivalently at the other boundary $C_{n+1}$), and using Eqs \ref{eqAsol} and \ref{bcampli},
one obtains
 \begin{eqnarray}
E_1 && \simeq  W(C_0 \to C_1)\left[1- \frac{A(C_1)}{ A(C_0)} \right] 
\nonumber \\ 
&& \simeq  W(C_0 \to C_1)\left[1- 
 \frac{R_0(1,n)}{R_0(0,n)} - \frac{A(C_{n+1})} { A(C_0)}\frac{R_{n+1}(1,1)}{R_{n+1}(1,n)}
 \right] 
\nonumber \\ 
&& \simeq  \frac{ W(C_0 \to C_1) }{R_{n+1}(1,n+1)} \left[1  - \frac{A(C_{n+1})} { A(C_0)} \right] 
\nonumber \\ 
&& \simeq  \frac{ W(C_0 \to C_1) }{R_{n+1}(1,n+1)} \left[1  + e^{ \beta \left[U(C_{n+1})- U(C_0)\right] } \right] 
\label{e1res}
\end{eqnarray}
Using Eq. \ref{defppm} and Eq. \ref{kestennp1}, one finally obtains
 \begin{eqnarray}
\frac{1}{E_1} &&  =  \frac{R_{n+1}(1,n+1)}
{ W(C_0 \to C_1)\left[1  + e^{ \beta \left[U(C_{n+1})- U(C_0)\right] } \right] } 
\nonumber \\ 
&& =  \frac{1}
{ W(C_0 \to C_1)\left[1  + e^{ \beta \left[U(C_{n+1})- U(C_0)\right] } \right] }
\left[  1+\sum_{m=1}^{n} \prod_{i=1}^{m} \frac{ p_-(i)}{ p_+(i)} \right]
\nonumber \\ 
&& =  \frac{1}
{ W(C_0 \to C_1)\left[1  + e^{ \beta \left[U(C_{n+1})- U(C_0)\right] } \right] }
\left[  1+\sum_{m=1}^{n} \prod_{i=1}^{m} \frac{W(C_i \to C_{i-1})}{W(C_i \to C_{i+1})} \right]
\label{e1invw}
\end{eqnarray}

\subsection{ Renormalized amplitude $G_R$ for  an effective one-dimensional dynamics }

The renormalized quantum Hamiltonian is given by the projection onto
the two lowest eigenstates $E_0=0$ and $E_1$
\begin{eqnarray}
H^{eff} \simeq E_1   \vert \psi_1 > < \psi_1 \vert 
\label{Heffprojdef}
\end{eqnarray}
where the first excited state $\vert \psi_1 >$ is given by Eq \ref{psi1orthog} near zero temperature
So Eq. \ref{Heffprojdef} becomes
\begin{eqnarray}
H^{eff} && \opsimeq_{\beta \to +\infty}  \frac{E_1}{  e^{- \beta U(C_0) } + e^{- \beta U(C_{n+1}) }}  
 [ e^{- \beta U(C_{n+1}) } \vert C_0 > < C_0 \vert
+ e^{- \frac{\beta}{2} U(C_0) }\vert C_{n+1} >  < C_{n+1} \vert 
\nonumber \\
&& \ \ \ \ \ \ \ \ \ \ \ \ \ \ \ \ \ \
 - e^{- \frac{\beta}{2} \left[ U(C_0)+U(C_{n+1}) \right]}  \left( \vert C_{0} >  < C_{n+1} \vert
+ \vert C_{n+1} >  < C_{0} \vert \right) ]
\label{Heffproj}
\end{eqnarray}
So it is of the form of Eq. \ref{tight} with only the two configurations $C_0$ and $C_{n+1}$
\begin{eqnarray}
H^{eff}   \opsimeq_{\beta \to +\infty}  &&   G_R({ C_0} , { C_{n+1}})
\label{tighteff}  \\
&& \left[  
e^{- \frac{\beta}{2} \left[  U({ C_{n+1}})- U({ C_0} \right] } \vert { C_0 } > < { C_0 } \vert
+ e^{- \frac{\beta}{2} \left[  U({ C_0})- U({ C_{n+1}} \right] } \vert { C_{n+1} } > < { C_{n+1} } \vert
 - \vert { C_0} > < { C_{n+1} } \vert -  \vert C_{n+1} >  < C_{0}\right]
\nonumber
\end{eqnarray}
where the renormalized amplitude reads
\begin{eqnarray}
G_R ({ C_0} , { C_{n+1}}) && \simeq \frac{E_1 e^{- \frac{\beta}{2} \left[ U(C_0)+U(C_{n+1}) \right]} }{  e^{- \beta U(C_0) } + e^{- \beta U(C_{n+1}) }}  = \frac{E_1 e^{ \frac{\beta}{2} \left[ U(C_0)+U(C_{n+1}) \right]} }{  e^{ \beta U(C_0) } + e^{ \beta U(C_{n+1}) }}
\label{Gr}
\end{eqnarray}

Using Eq. \ref{e1invw}, the final formula for the renormalized amplitude reads
\begin{eqnarray}
\frac{1}{G_R ({ C_0} , { C_{n+1}})} && =  \frac{ 2 \cosh \frac{\beta}{2} \left[ U(C_0)-U(C_{n+1}) \right]  } {E_1  } 
\nonumber \\
&& =  \frac{ e^{ \frac{\beta}{2} \left[U(C_{0})- U(C_{n+1} )\right] } }
{ W(C_0 \to C_1) }
\left[  1+\sum_{m=1}^{n} \prod_{i=1}^{m} \frac{W(C_i \to C_{i-1})}{W(C_i \to C_{i+1})} \right]
\label{Grfinalw}
\end{eqnarray}
This formula is used to obtain Eq. \ref{Grfinalwbord} and Eq. \ref{Grfinalwbulk} of the text.

\subsection{ Example of application }

Let us now consider the case where the transition rates $W(C \to C')$ of the effective one-dimensional
problem of Eqs \ref{eqAgenebulk} and \ref{eqAgenebord} satisfy the detailed balance form of Eq. \ref{gdetailed}.
Then we may rewrite the products as
\begin{eqnarray}
 \prod_{i=1}^{m} \frac{W(C_i \to C_{i-1})}{W(C_i \to C_{i+1})}
&&  = \prod_{i=1}^{m} 
\left( \frac{  G\left(  C_i ,  C_{i-1}  \right) e^{- \frac{\beta}{2} \left[  U({ C_{i-1}})- U({ C_i} \right] }  }
{ G\left(  C_i ,  C_{i+1} \right) e^{- \frac{\beta}{2} \left[  U({ C_{i+1}})- U({ C_i} \right] } } \right)
\nonumber \\ 
&& =  \frac{  G\left(  C_0 ,  C_{1}  \right)  }
{ G\left(  C_m ,  C_{m+1} \right)  }
 e^{ \frac{\beta}{2} \left[  U({ C_{m}})+U({ C_{m+1}})- U({ C_0}) - U({ C_1})  \right] } 
\label{prodppm}
\end{eqnarray}
so that the renormalized amplitude of Eq. \ref{Grfinalw} reads 
\begin{eqnarray}
\frac{1}{G_R ({ C_0} , { C_{n+1}})}
&& =  \frac{ e^{ \frac{\beta}{2} \left[U(C_{0})- U(C_{n+1} )\right] } }
{ G(C_0 , C_1)e^{- \frac{\beta}{2} \left[  U({ C_{1}})- U({ C_0} \right] } }
\left[  1+\sum_{m=1}^{n}  \frac{  G\left(  C_0 ,  C_{1}  \right)  }
{ G\left(  C_m ,  C_{m+1} \right)  }
 e^{ \frac{\beta}{2} \left[  U({ C_{m}})+U({ C_{m+1}})- U({ C_0}) - U({ C_1})  \right] }   \right]
\nonumber \\ 
&& = e^{ - \frac{\beta}{2} \left[ U(C_0)+U(C_{n+1}) \right]}
\sum_{m=0}^{n} \frac{    e^{ \frac{\beta}{2} \left[  U({ C_{m}})+U({ C_{m+1}})  \right] } }
{ G\left(  C_m ,  C_{m+1} \right)  }
\label{Grfinal}
\end{eqnarray}
This formula is used to obtain Eq. \ref{Grbord} and Eq. \ref{Grbulk} of the text.

\end{document}